\documentclass[preprint2]{aastex}

\usepackage{natbib,amssymb}
\usepackage{color}
\usepackage[normalem]{ulem}

\newcommand{\amm}{NH$_3$}
\newcommand{\dia}{N$_2$H$^+$}

\newcommand{\hcn}{HC$_7$N}

\newcommand{\kms}{km\,s$^{-1}$}
\newcommand{\cc}{cm$^{-3}$}
\newcommand{\vlsr}{$v_{\mathrm{LSR}}$}
\newcommand{\tk}{$T_\mathrm{K}$}

\newcommand{\nh}{N(H$_2$)}
\newcommand{\mvir}{$M_\mathrm{vir}$}

\newcommand{\papi}{Paper I}
\renewcommand{\S}{Section} 


\shorttitle{}
\shortauthors{Friesen et al.}
\shorttitle{Hierarchical fragmentation in Serpens South}

\def\edits#1{{{#1}}}
\def\referee#1{{{#1}}}
\slugcomment{Draft version \today,  Accepted to The Astrophysical Journal}

\begin{document}

\title{The fragmentation and stability of hierarchical structure in Serpens South}
\author{R. K. Friesen\altaffilmark{1}}
\altaffiltext{1}{Dunlap Institute for Astronomy \& Astrophysics, University of Toronto, Toronto, Ontario, Canada, M5S 3H4}
\email{friesen@dunlap.utoronto.ca}
\author{T. L. Bourke\altaffilmark{2}}
\altaffiltext{2}{SKA Organisation, Jodrell Bank Observatory, Lower Withington, Macclesfield SK11 9DL, UK}
\author{J. Di Francesco\altaffilmark{3,4}}
\altaffiltext{3}{National Research Council Canada, Radio Astronomy Program, 5071 West Saanich Rd, Victoria, BC, V9E 2E7, Canada}
\altaffiltext{4}{Department of Physics and Astronomy, University of Victoria, Victoria, BC, V8P 1A1, Canada}
\author{R. Gutermuth\altaffilmark{5}}
\altaffiltext{5}{Department of Astronomy, University of Massachusetts Amherst, Amherst, MA 01003}
\author{P. C. Myers\altaffilmark{6}}
\altaffiltext{6}{Radio and Geoastronomy Division, Harvard Smithsonian Center for Astrophysics, MS-42, Cambridge, MA, 02138, USA}

\begin{abstract}

Filamentary structures are ubiquitous in molecular clouds, and have been recently argued to play an important role in regulating the size and mass of embedded clumps through fragmentation and mass accretion. Here, we reveal the dynamical state and fragmentation of filamentary molecular gas associated with the Serpens South protocluster through analysis of wide ($\sim 4\ \mathrm{pc} \times \ 4\ \mathrm{pc}$) observations of \amm\ (1,1) and (2,2) inversion transitions with the Green Bank Telescope. Detailed modeling of the \amm\ lines reveals that the kinematics of the cluster and surrounding filaments are complex. We identify hierarchical structure using a dendrogram analysis of the \amm\ emission. The distance between neighbour structures that are embedded within the same parent structure is generally greater than expected from a spherical Jeans analysis, and is in better agreement with cylindrical fragmentation models. The \amm\ line width-size relation is flat, and average gas motions are sub- or trans-sonic over all physical scales observed. Subsonic regions extend far beyond the typical 0.1\ pc scale previously identified in star-forming cores. As a result, we find a strong trend of decreasing virial parameter with increasing structure mass in Serpens South. Extremely low virial parameters on the largest scales probed by our data suggest that the previously observed, ordered magnetic field is insufficient to support the region against collapse, in agreement with large radial infall motions previously measured toward some of the filaments. A more complex magnetic field configuration in the dense gas, however, may be able to support the filaments.

\end{abstract}


\section{Introduction}
\label{sec:intro}

Most stars in our galaxy do not form in isolation \citep{lada03}, but instead form in groups and clusters. Many clustered star-forming regions share similar morphologies, where the greatest star formation rates are found within a central `hub' of dense molecular gas, while streams or filaments of additional material (still containing substantial mass) extend away from the hub \citep[e.g., ][]{myers09}. In nearby molecular clouds, much of the dense gas that is critical to star formation is found in filaments \citep{andre14, konyves15}. Filamentary structures may thus play a key role in regulating the fragmentation of dense gas into clumps and cores. Here, we investigate the structure, stability and fragmentation of dense molecular gas in the young, stellar-cluster forming region, Serpens South. 

Serpens South contains a bright protostellar cluster embedded within a hub-filament type structure of dense gas traced by Spitzer 8~\micron\, absorption \citep{gutermuth08} and dust continuum emission \citep{andre10,maury11}. \citeauthor{maury11} identify 57 protostellar objects with $M > 0.1 \ \mathrm{M}_\odot$ within the central cluster, confirming the high star formation rate and young cluster age first determined by \citeauthor{gutermuth08} The associated filaments, which have a significantly lower star formation rate than the central cluster, contain $M \sim 610~\mathrm{M}_\odot$ of molecular gas and dust (where Maury et al. assume a distance to the cluster $d = 260$~pc; at a revised distance of 429\ pc, discussed further below, $M \sim 1660\ \mathrm{M}_\odot$). Serpens South is thus a young, active star-forming region that contains sufficient mass to form a substantial number of new stars in the future, and is an ideal target for studies on the dynamic importance of filamentary structure in star-forming regions. 

\begin{figure*}
\includegraphics[trim=15 15 70 120, clip,width=\textwidth]{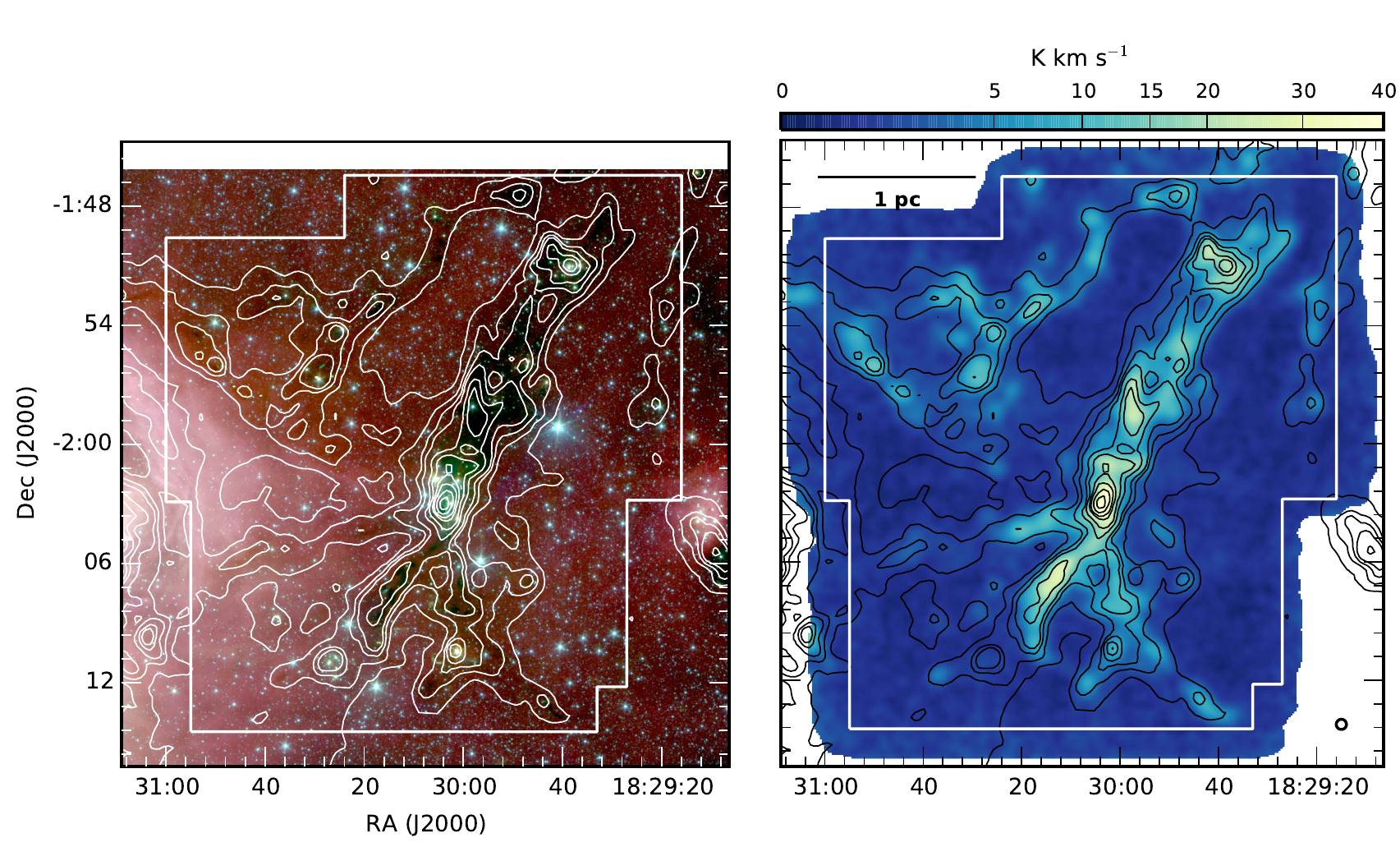}
\caption{Left: Spitzer RGB image (8\ \micron, 4.5\ \micron, 3.6\ \micron) of Serpens South overlaid with \edits{500\ \micron\, continuum contours at 36\arcsec\ resolution \citep[HPBW;][]{konyves15}. Contour levels are 4, 6, 8, 10, 15, 20, 30, 40, 60, 90 $\times$ 16\ MJy\ sr$^{-1}$.} Right: Integrated \amm\ (1,1) intensity in K~\kms\, observed with the GBT KFPA at 32\arcsec\ resolution (FWHM). In both panels, the white outline shows the approximate area where the rms noise in the GBT KFPA \amm\ (1,1) cube is $\lesssim 0.1$~K. \label{fig:mm11}}
\end{figure*}

Figure \ref{fig:mm11} (left) presents a Spitzer three-color image (8~\micron, 4.5~\micron, 3.6~\micron) of the Serpens South star-forming region \citep{gutermuth08}, overlaid with dust continuum emission contours at 500~\micron\, observed at 36\arcsec\, resolution (HPBW) with the Herschel Space Observatory as part of the Herschel Gould Belt Survey \citep{andre10,konyves15}. The central star-forming hub of Serpens South is visible near the image centre, while dark lanes show high column density regions revealed by 8~\micron\, absorption in the Spitzer data. The continuum contours follow well the 8~\micron\, absorption features, and furthermore reveal dense molecular clumps and cores embedded within the large filaments that extend north, east, and south from the central cluster. Bright infrared emission to the east originates from the W40 HII region. 

Toward Serpens South, the central cluster's protostars have been studied in the infrared and radio \citep{gutermuth08,kern16}. Analysis of the outflows originating from the protostellar cluster shows that the energy injection rate by the outflows into the gas is likely sufficient to maintain supersonic turbulence in the region, but cannot disperse the clump of molecular gas associated with the cluster \citep{nakamura11,plunkett15}. Narrow filaments are seen in the dense gas near the central cluster \citep{flopez14}. On larger scales, \citet{kirk13} argued that at least one filament shows evidence for material flow onto the central cluster, with a resulting mass accretion rate high enough to sustain the current star formation rate. In a companion paper to this one, we found evidence for accretion of low density material onto the filaments through analysis of the distribution and kinematics of the emission from the cyanopolyyne \hcn\ \citep[][hereafter Paper I]{friesen13}. The accretion rate onto one of the filaments is sufficient to double the filament mass in $\sim 1-2$~Myr, similar to the typical star formation timescale. \citet{tanaka13} show that on $\sim 0.1$~pc scales, the dense filamentary gas near the central cluster has low virial parameters of only $\sim 0.1-0.3$, suggesting that forms of support other than gas turbulence are needed to prevent free-fall collapse. The magnetic field in the low density gas surrounding the cluster and dense gas complex is well-ordered, with a magnetic field direction largely perpendicular (in projection) to the main north-south filaments extending from the central cluster \citep{sugitani11}. 

\citeauthor{gutermuth08} and several subsequent papers, including \papi, assumed a distance of $\sim 260$\ pc toward Serpens South, matching the photometry-derived distance to Serpens Main \citep{straizys03}, given the similar local standard of rest velocities (\vlsr) of the two regions and their near locations within the Aquila ridge. Further discussion of the multiple distance estimates toward Serpens Main, W40, and Serpens South is given in \citet{maury11} and \citet{plunkett15}. \referee{Very Long Baseline Array trigonometric parallax measurements toward several young stars associated with Serpens Main and the W40 complex, however, have shown conclusively that these regions are instead at a greater distance of $\sim 430$\ pc \citep[$429\pm2$\ pc and $439\pm9.2$\ pc, respectively;][]{dzib11,ortiz-leon16}.} In this paper, we assume the \citeauthor{dzib11} distance of $429\pm2$\ pc for Serpens South, and note where this assumption results in conflict with other published studies that assume the nearby distance of 260\ pc. 

Here, we investigate the structure, stability, and fragmentation of dense molecular gas over most of the filamentary complex in Serpens South. We present Green Bank Telescope (GBT) observations of the \amm\ (1,1), (2,2) and (3,3) inversion transitions toward a 30\arcmin\ $\times$ 31\arcmin\ area ($3.7\ \mathrm{pc} \times 3.9\ \mathrm{pc}$ at $d = 429$\ pc) around the Serpens South cluster and associated molecular gas. \amm\ is an excellent tracer of dense ($n \gtrsim 10^4$\ \cc), quiescent gas, and observations of multiple inversion transitions allow direct measurement of the kinetic gas temperature \citep{hotownes}. In \S\ \ref{sec:obs}, we briefly describe the observations and data reduction, but refer the reader to \papi\ for a more detailed discussion. In \S\ \ref{sec:results}, we present the resulting integrated intensity maps and the gas property maps determined through hyperfine modeling of the \amm\ lines. We furthermore introduce the dendrogram analysis used on the 3D data cubes to identify hierarchical structure in the \amm\ emission. In \S\ \ref{sec:discussion}, we analyse the stability and fragmentation hierarchical structure of dense gas in Serpens South. We show that structures in Serpens South are virially unstable in the absence of strong magnetic field support, in agreement with previous measurements of infall toward the main gas filaments surrounding the protocluster. We also find that fragmentation in Serpens South is in better agreement with cylindrical fragmentation models than predictions from a spherical Jeans analysis. In \S\ \ref{sec:summary}, we provide a summary of our key results. 

\section{Observations}
\label{sec:obs}

Observations of the \amm\ $(J,K) = (1,1), (2,2), (3,3)$ and \hcn\, $J=21-20$ emission lines (rest frequencies of 23.694~GHz, 23.723~GHz, 23.870~GHz, and 23.68789~GHz, respectively) were made toward Serpens South using the 7-element K-band Focal Plane Array (KFPA) at the Green Bank Telescope (GBT) between November 2010 and April 2011 in shared risk time. For all lines, the total Nyquist-sampled map extent is approximately $30\arcmin\, \times 31\arcmin$, while the angular resolution is 32\arcsec\, (FWHM), or 0.07~pc. Details of the observations and data reduction using the GBT KFPA data reduction pipeline \citep{masters11} are presented in \papi, along with detailed analysis of the \hcn\ data. The mean rms noise in the off-line channels of the \amm\ (1,1) and (2,2) data cubes is 0.06\,K in 0.15~\kms\, channels,  with higher values ($\sim 0.1$\,K) near the map edges where fewer beams overlap. In general, the noise in the map is constant, with a 1-$\sigma$ variation in the rms noise of only 0.01\,K in the region where all the KFPA beams sample the sky. The white outline in Figure \ref{fig:mm11} identifies the region observed by the GBT to a sensitivity of $\lesssim 0.1$~K rms. The \amm\ (3,3) map, subtending the same area but mapped with a single beam, has a mean rms noise level of 0.09\,K, with greater variation between sub-maps. Data cubes, moment maps and property maps (described below) are publicly available  \citep{friesen16data}. The data presented in \papi\ are also publicly available \citep{friesen13data}.  

\section{Results and Analysis}
\label{sec:results}

\subsection{Maps}

\begin{figure}
\includegraphics[width=0.48\textwidth]{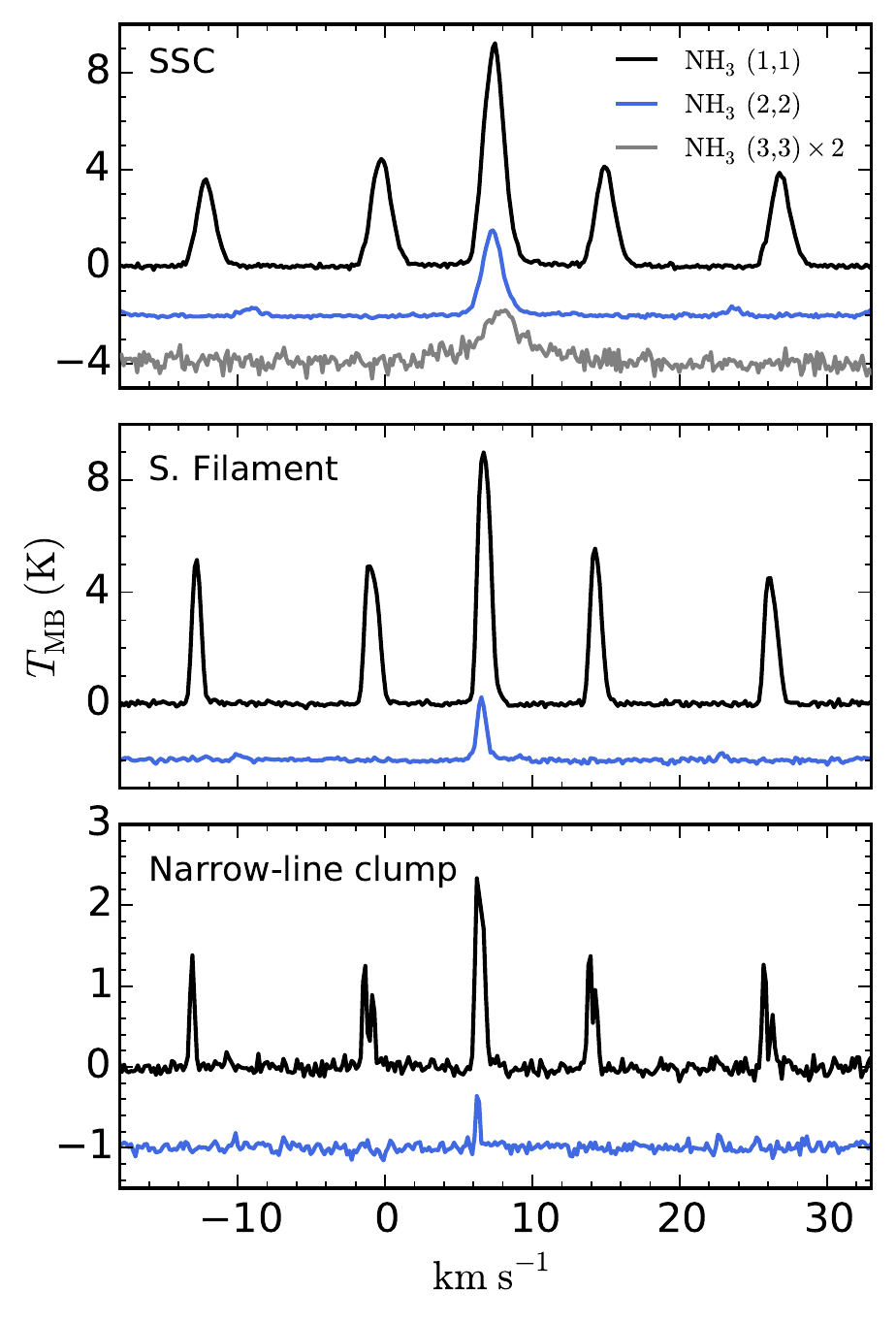}
\caption{\amm\ (1,1), (2,2), and (3,3) spectra (where detected) toward the \amm\ emission peak in the Serpens South protocluster (top), the location of brightest emission in the filament extending south-east from the cluster (middle), and a narrow-line clump southwest of the cluster (bottom). \amm\ (3,3) emission (here multiplied by a factor of two) was detected only toward the cluster, and shows both a $v_\mathrm{LSR}$ offset relative to the (1,1) and (2,2) emission, as well as broad line wings unseen in the lower order transitions. \label{fig:spectra}}
\end{figure}

\begin{figure}
\includegraphics[width=0.48\textwidth]{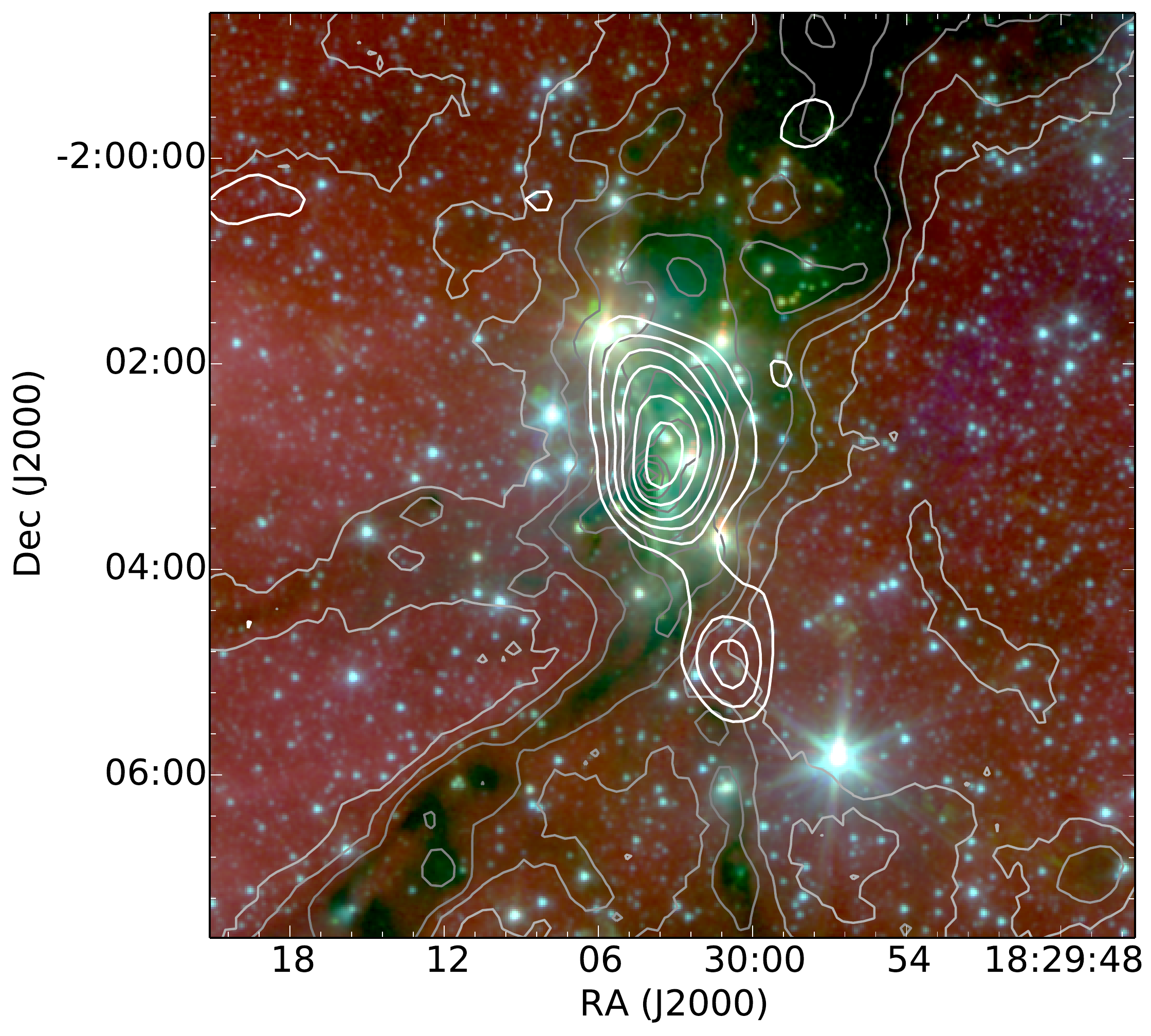}
\caption{Spitzer RGB image toward the central Serpens South cluster only, overlaid with \amm\ (3,3) integrated intensity contours (white; 3, 5, 7, 9, 12, 15$ \times \sigma$) and 850\ \micron\, continuum contours (grey). RGB image and continuum contours as in Figure \ref{fig:mm11}. \amm\ (3,3) emission was seen only toward the central YSO cluster.  \label{fig:33m0}}
\end{figure}

Figure \ref{fig:mm11} shows a comparison of infrared and \amm\ (1,1) zeroth moment emission toward Serpens South, overlaid with 500\ \micron\ contours. In general, \amm\ (1,1) zeroth moment emission follows very well the 850~\micron\, continuum contours, with some exceptions toward large but low-brightness continuum features (with consequently lower column density), and as well as toward some bright continuum peaks that are coincident with point sources (likely protostars) in the infrared data. Furthermore, several features are visible in the \amm\ (1,1) integrated intensity map that are less prominent in the continuum data. To our sensitivity, \amm\ (1,1) is detected over most of the map. \referee{We find strong line emission toward the central cluster and connected filaments, as well as toward the more fragmented structures in the north-east. Fainter emission is also present in the regions between these \amm-bright features. In general, the \amm\ line widths are more broad toward the central cluster, as well as in the lower brightness emission between filaments and cores. The filaments and cores themselves typically have more narrow line emission. Examples of \amm\ line profiles in the central cluster, filament, and isolated core are shown in Figure \ref{fig:spectra}, while the \amm\ line widths will be discussed in more detail below.} \amm\ (2,2), not shown, is well-detected over most of the main Serpens South filaments and clumps and closely follows the \amm\ (1,1) emission, but with lower signal-to-noise ratio (SNR). Figure \ref{fig:33m0} shows \amm\ (3,3) zeroth moment contours overlaid on the infrared image of the central cluster.  \amm\ (3,3) is detected only toward the central cluster, and is clearly kinematically distinct from the \amm\ (1,1) and (2,2) emission, with larger line widths and extended line wings suggestive of excitation in outflows, and line-of-sight velocity shifts up to $\sim 1$\ \kms\ relative to the (1,1) and (2,2) lines. This behavior is seen in Figure \ref{fig:spectra}, where we show a selection of \amm\ spectra toward the central cluster and two other locations in Serpens South.

\subsection{Line fitting}
\label{sec:line_fit}

The \amm\ molecule exists in two forms, ortho-\amm\ and para-\amm, depending on the orientation of the three identical hydrogen nuclear spins. Transitions between ortho- and para-\amm\ are forbidden, since radiative and non-reactive collisional transitions do not normally change the spin orientations. Ortho- and para-\amm\ must be therefore treated as separate species. The ratio of ortho- to para-\amm\ depends on the temperature at which the \amm\ formed. At formation temperatures $\gtrsim 40$\ K, the ortho- to para-\amm\ ratio is approximately 1, increasing to $\sim 4$ at $T \sim 10$~K, since the (0,0) level is an ortho-\amm\ level \citep{takano02}. The (1,1) and (2,2) inversion transitions belong to para-\amm, while (3,3) is an ortho-\amm\ transition. Since we only detect (3,3) emission toward the cluster center, where it shows a significantly different distribution than that of (1,1) and (2,2), we cannot place limits on the ortho- to para-\amm\ ratio in Serpens South, and in the following analysis we focus only on the (1,1) and (2,2) transitions. 

We fit simultaneously each observed \amm\ (1,1) and (2,2) spectrum with a custom Gaussian hyperfine structure fitting routine written in IDL and described in \papi. The routine creates model spectra given input kinetic gas temperature, \tk, line-of-sight velocity relative to the local standard of rest (LSR), \vlsr, line full-width at half maximum, $\Delta v$, \amm\ (1,1) opacity, $\tau_{(1,1)}$, and excitation temperature, $T_{ex}$, and performs a chi-square minimization against the observed \amm\ emission using the {\sc idl} {\sc mpfit} routines \citep{markwardt}. The routine assumes a two-level system, with common \vlsr, $\Delta v$, and $T_{ex}$ for the (1,1) and (2,2) hyperfine components. 

\begin{figure*}
\begin{center}
\includegraphics[trim=40 0 140 0, clip, width=0.98\textwidth]{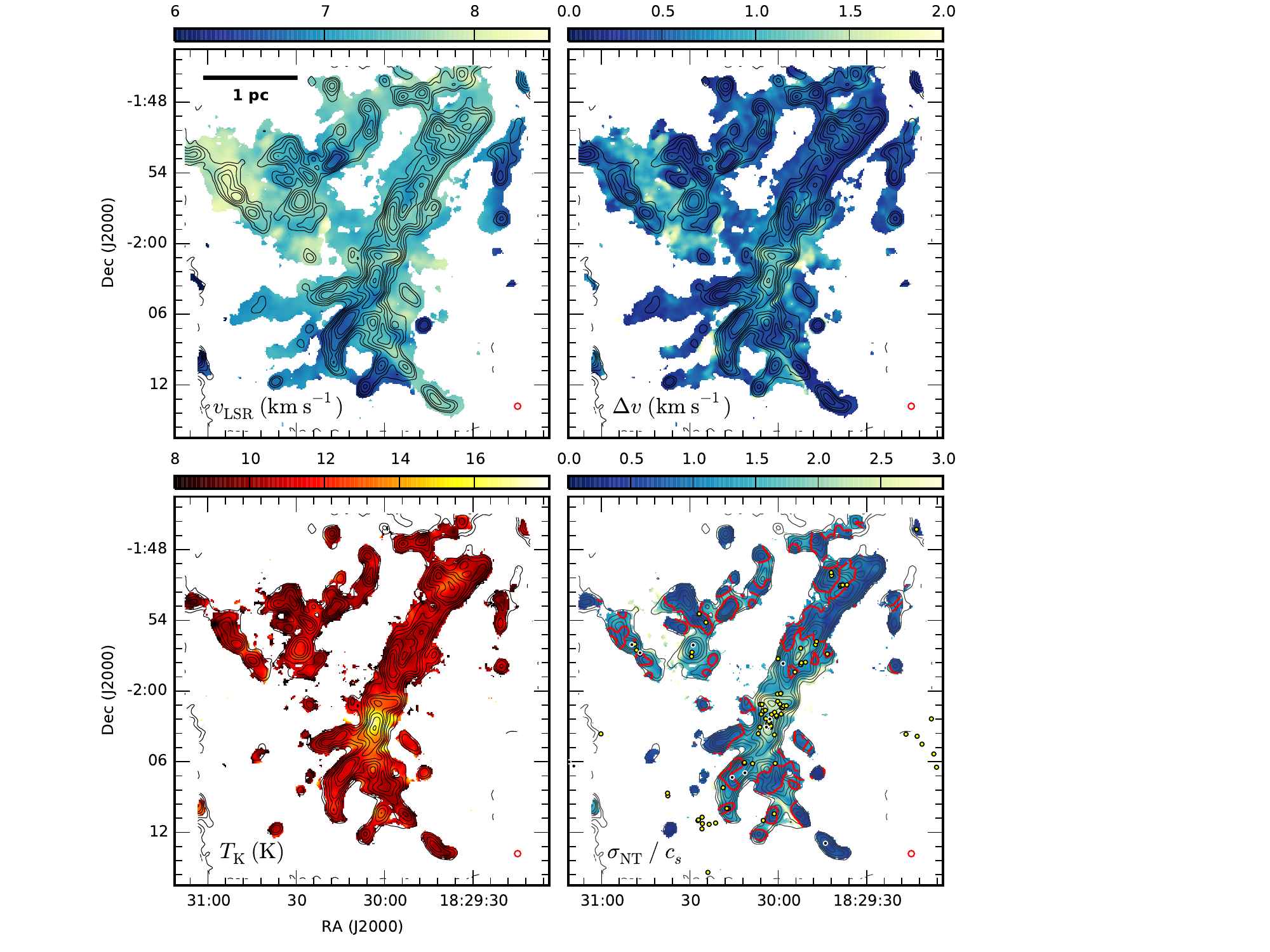}
\caption{Top left: \vlsr\ (\kms) toward Serpens South from \amm\ (1,1) and (2,2) hyperfine line fitting. \edits{In all plots, black contours show integrated \amm\ (1,1) intensity contours at 2, 3, 5, 7.5, 10, 15, 20, 30, 40\ K\ km\ s$^{-1}$.} The 32\arcsec\ GBT beam at 23.7\ GHz is shown by the red circle at bottom right. Top right: $\Delta v$ (\kms) toward Serpens South. Lower left: Gas temperature \tk\ (K), masked where insufficient SNR in the \amm\ (2,2) line precluded a good fit. Lower right: The ratio of the non-thermal velocity dispersion, $\sigma_\mathrm{NT}$, to the thermal sound speed, $c_s$, derived from the \amm\ line width and \tk\ measurements, masked over the same region as \tk. The red contour highlights the transition from super- or trans-sonic to subsonic velocity dispersions ($\sigma_\mathrm{NT} / c_s = 1$). Black and yellow circles show the locations of embedded Class 0 and Class I protostars, respectively \citep{gutermuth09}. 
\label{fig:linefits}}
\end{center}
\end{figure*}

In the first three panels of Figure \ref{fig:linefits}, we show the resulting maps of \vlsr, $\Delta v$, and \tk. We apply more stringent masking to the \tk\ map since accurate fits to these parameters require an SNR $\gtrsim 3$ in the \amm\ (2,2), while \vlsr\ and $\Delta v$ can be determined from the \amm\ (1,1) line without a matching detection in \amm\ (2,2). \referee{To account for artificial broadening of the reported line widths by the 0.15\ \kms\ velocity resolution, we have subtracted in quadrature the resolution width in the discussion of $\Delta v$ below and in the rest of the paper.}

Figure \ref{fig:linefits} shows that there is a range of $\sim 2$~\kms\ in \vlsr\ across Serpens South, with lower velocity features primarily associated with several filaments to the southeast and northwest. One exception is the presence of an elongated core or filament toward the northeast with a dramatic shift of $\sim 0.5$~\kms\, in \vlsr\ relative to the surrounding gas. The bulk of the dense gas has $v_\mathrm{lsr} \sim 7-7.5$~\kms, while higher velocities are found in a ridge that connects the gas in the northeast near W40 with the most extended, narrow filament in the southwest, as well as in scattered clumps around the main north-south filament. 

Most of the filamentary and clumpy features present in the continuum and \amm\ integrated intensity maps are associated with relatively narrow \amm\ line widths, with $\Delta v \sim 0.5$~\kms. Toward some regions, the \amm\ line widths are consistent with being nearly purely thermal within our velocity resolution \referee{of 0.15\ \kms. Given the hyperfine structure of the \amm\ lines, we are able to fit well line widths that are near the velocity resolution of our data where the emission is bright. For example, the fit to the spectrum shown in Figure \ref{fig:spectra} for the narrow line clump gives a line width $\Delta v = 0.21 \pm 0.005$\ \kms, or $\sigma_v = \Delta v / 2.3548 = 0.09$\ \kms, while the thermal velocity dispersion for \amm\ is $0.07$\ \kms\ at a temperature of 10\ K. Since the observed line width is similar to our velocity resolution, the fit value may overestimate the true velocity dispersion of the clump.} Outside of the filaments, where we still have sufficiently high SNRs, and near the central cluster, however, we see large changes in the observed line width. In several regions, line widths transition sharply between wide and narrow values in less than a beam width (0.07~pc at $d = 429$~pc). In some regions,  changes in $\Delta v$ are correlated with shifts in \vlsr. Given that our \amm\ fitting routine uses a single velocity component only, an \amm\ line resulting from two gas components at slightly different velocity overlapping along the line-of-sight would manifest as a \referee{region of increased line width and velocity intermediate to the two components. Such increased line widths can be seen along the edges of filaments, for example. Where an \amm-bright clump is surrounded by emission at a different velocity and with lower brightness temperature, the routine fits a broader line width and intermediate velocity in the region where the emission from the clump is increasing but is still comparable in line brightness to the surrounding material, resulting in a} ring of large fitted $\Delta v$ around a region of smaller values. Such rings are indeed visible toward some parts of the map. In other cases, the sharp $\Delta v$ transition is present without a corresponding shift in \vlsr. We relate these transitions to the thermal sound speed of the gas, and discuss further, below. 

While we are able to fit \vlsr\ and $\Delta v$ over much of Serpens South, we are only able to determine well \tk\ along the main filaments and clumps where the SNR of the \amm\ (2,2) emission is greater. Within these structures, the gas temperature remains remarkably constant, with a mean value of 11\ K and a standard deviation of only $\sim 1$\ K, as shown in the third panel of Figure \ref{fig:linefits}. Heating can be seen toward the central YSO cluster, where the gas temperatures rise to a peak of 18\ K. Moving away from the central cluster along the main filament, \tk\ declines roughly with $1/r^2$, reaching the mean filament temperature of $\sim 11$~K at a distance of $\sim 0.5$~pc (4\arcmin) from the cluster centre. Away from the central cluster, localized heating (only $\sim 1-2$\ K in most cases) is often, but not always, associated with smaller clusters of protostars, or at lower \amm\ integrated intensity contours where the gas density is likely lower. 

Overall, the gas temperatures in Serpens South are similar to those found in other dark filaments, such as Infrared Dark Clouds \citep[IRDCs;][]{ragan11} and cores in nearby molecular clouds \citep[e.g.,][]{rosolowsky08}, and are slightly less than the typical gas temperatures at similar spatial resolution in nearby cluster-forming regions like Orion \citep{li13} and Ophiuchus \citep{friesen09}, and in dense molecular gas within high mass star-forming regions \citep[$\gtrsim 20$\ K;][]{molinari96,hill10}. On average, the \amm-derived kinetic gas temperatures are lower than the dust temperatures derived from spectral energy distribution fitting at submillimetre wavelengths \citep{konyves15}. This difference may be due to \amm\ being selectively excited at higher densities in colder, starless regions, whereas the continuum emission includes contributions from all dust along the line-of-sight. A detailed comparison between \amm\ and dust temperatures is beyond the scope of this paper. 

\edits{From the \amm\ line width $\Delta v$ and kinetic temperature \tk, we determine the non-thermal velocity dispersion of the \amm\ emission, 
$\sigma_\mathrm{NT} = \sqrt{\sigma_v^2 - k_\mathrm{B} T_\mathrm{K} / m_\mathrm{NH3}}$, 
where $k_\mathrm{B}$ is the Boltzmann constant, $m_\mathrm{NH3}$ is the mass of an \amm\ molecule, and 
$\sigma_v = \Delta v / (8 \ln 2)^{-1/2}$ 
is the velocity dispersion of the line. We evaluate $\sigma_\mathrm{NT}$ only where we have well-fit $T_\mathrm{K}$ values. The last panel in Figure \ref{fig:linefits} shows the ratio of $\sigma_\mathrm{NT}$ to the thermal sound speed, $c_s = \sqrt{k_\mathrm{B} T / \mu m_\mathrm{H}}$, where $m_\mathrm{H}$ is the mass of a hydrogen atom and $\mu = 2.33$ is the mean molecular weight per free particle. We find that much of the dense gas in Serpens South has subsonic or trans-sonic non-thermal motions, where the mean $\sigma_\mathrm{NT}$ across the region is similar to the expected $\sim 0.2$\ \kms\ sound speed at 11\ K. Toward multiple locations, however, $\sigma_\mathrm{NT} \lesssim 0.1$\ \kms, reaching the limit of our spectral resolution\referee{. Here, the non-thermal velocity dispersion is substantially smaller than the sound speed, and is comparable to the thermal component for \amm\ (0.07\ \kms\ at 11\ K). }}

In Figure \ref{fig:linefits}, we highlight in red the transsonic transition, i.e., where $\sigma_\mathrm{NT} / c_s = 1$, and furthermore show the locations of Class 0 and Class I protostars \citep{gutermuth09}. \referee{The Figure shows that subsonic gas motions are primarily associated with \emph{starless} peaks or filamentary structures in the integrated \amm\ emission. In some cases, rapid transitions between nearly thermal to supersonic occur in less than a beam width (0.07\ pc). In several regions, these transitions are correlated with rapid changes in \vlsr, as expected in a complex region like Serpens South, where structures may overlap along the line-of-sight. One example is seen at the apparent intersection of two filamentary structures at different \vlsr\ in the south-west. In other areas, such as within the filament extending north from the central protocluster, the transition from transsonic to subsonic gas occurs without a corresponding change in \vlsr. In both cases, the Figure shows that subsonic motions are prevalent over significantly larger spatial scales than typically identified in smaller-scale maps of dense gas within star-forming cores and filaments \citep[$\sim 1$\ pc versus $0.1$\ pc, e.g.,][]{pineda10,goodman98}. Similar subsonic filaments of lengths up to $\sim 6$\ pc have also been identified in Taurus and Musca \citep{tafalla15,hacar16}, but are associated with little ongoing star formation, and do not yet have sufficient mass to form an intermediate-mass stellar cluster like Serpens South. } 

\edits{Figure \ref{fig:linefits} shows that the increase in line widths near the central cluster is not due solely to the detected increase in temperature. Non-thermal motions increase out to some distance from the cluster centre. \referee{Using high-resolution CO observations of the many protostellar outflows emanating from the protocluster, \citet{plunkett15} show that the ratio of the outflow luminosity to the turbulent energy dissipation rate for the cluster (assuming similar non-thermal velocity dispersions as found here) implies that the turbulent energy in the region is consistent with being outflow-driven.} Furthermore, many other regions with greater non-thermal motions are associated with smaller groups of protostars, suggestive of an impact by their respective outflows on the dense gas motions\referee{, although detailed studies of the energy injection by outflows outside the cluster center do not yet exist.} Alternatively, \referee{or in conjunction with the injection of energy by outflows, some of} these more evolved regions may be undergoing large-scale gravitational collapse, such that the increased non-thermal motions are instead due to infall motions rather than turbulence. We discuss further the stability of structures seen in \amm\ emission in \S \ \ref{sec:stability}.}

\begin{figure}
\includegraphics[width=0.49\textwidth]{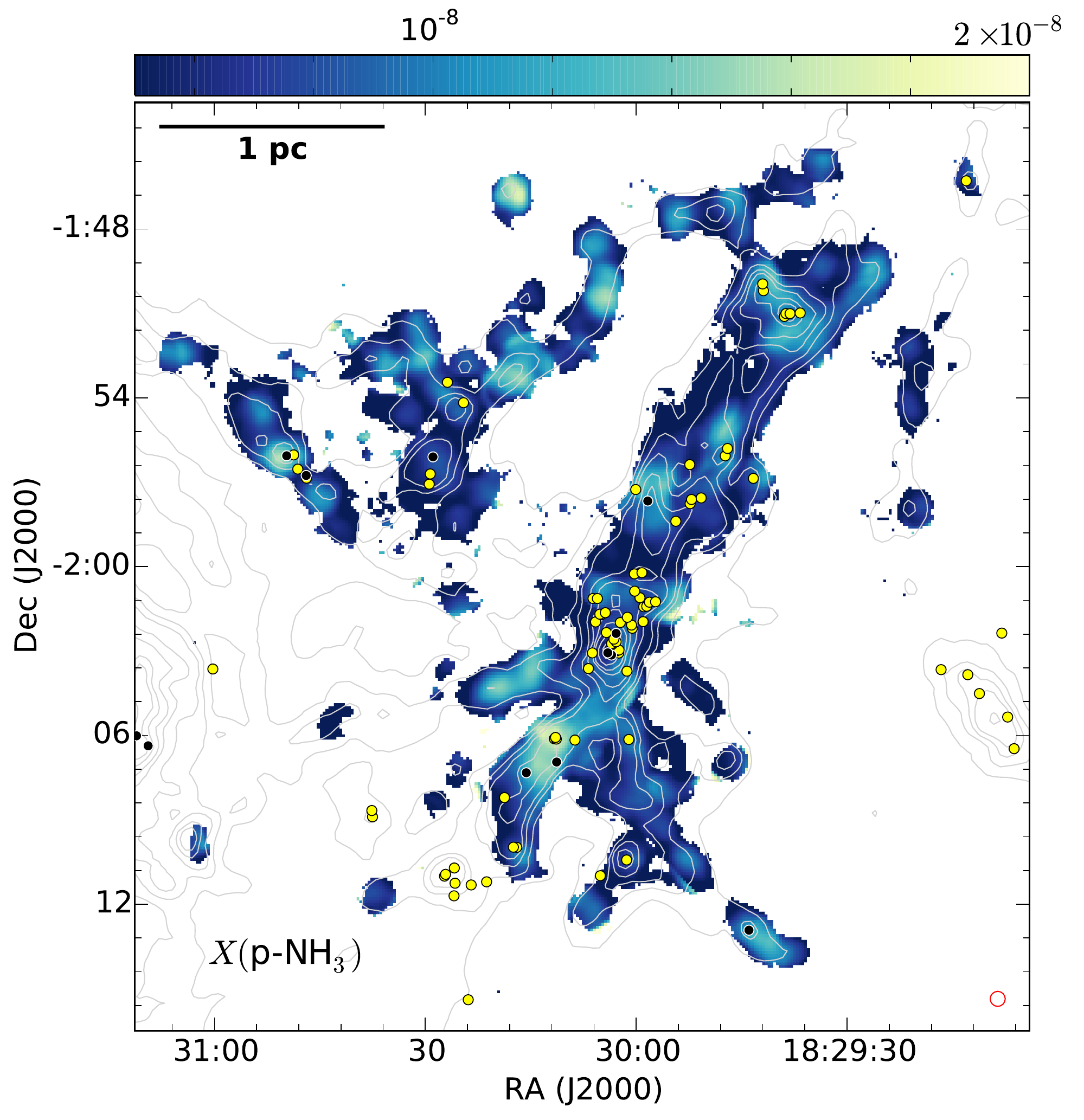}
\caption{The abundance of para-\amm\ relative to \nh. Contours show 500~\micron\, continuum emission as in Figure \ref{fig:mm11}. Black and yellow circles show the locations of embedded Class 0 and Class I protostars, respectively \citep{gutermuth09}. The 32\arcsec\ GBT beam at 23.7\ GHz is shown by the red circle at bottom right. \label{fig:xnh3}}
\end{figure}

\edits{We further calculate the column density of para-\amm\ following \citet{friesen09}, omitting the ortho- states in the partition function, and the para-\amm\ abundance relative to \nh\ using the H$_2$ column densities in \citet{konyves15}. Shown in Figure \ref{fig:xnh3}, we find a factor of several variation in the \amm\ abundance across Serpens South, with the lowest detected abundances, $X(\mathrm{NH}_3) \sim 5 \times 10^{-9}$, toward the central protostellar cluster, and the edges of clumps and filaments. Peak para-\amm\ abundances reach $X(\mathrm{NH}_3) \sim 2 \times 10^{-8}$. Also in Figure \ref{fig:xnh3}, the locations of Class 0 and Class I protostars are highlighted \citep{gutermuth09}. Frequently, clusters of Class I protostars correlate with lower \amm\ abundances, or a complete lack of \amm\ emission where continuum emission is still seen. Measurements of the total \amm\ abundance in nearby star-forming regions vary between $7 \times 10^{-10}$ \citep[B68; ][]{difran02} and a few times $10^{-8}$ \citep{hotzel01,tafalla06,crapsi07,friesen09}, with evidence for both constant and decreasing abundance at higher H$_2$ column densities in dense cores \citep[][respectively]{tafalla06,friesen09}. In infrared dark clouds, perhaps better analogues to the Serpens South complex, mean abundances can be as much as a few times $10^{-7}$ \citep{ragan11}. The para-\amm\ abundances calculated here are consistent with higher end of these ortho- plus para-\amm\ abundances, given typical values estimated for the ortho- to para-\amm\ ratio (resulting in $X(\mathrm{NH}_3) \sim 2 \times X(\mathrm{para-NH}_3)$). }

\subsection{Identifying \amm\ structures using dendrograms}
\label{sec:dendro}

Ammonia emission preferentially highlights cold ($T \sim 10-25$~K), dense ($n \gtrsim 10^4$~\cc) regions within molecular clouds. It is therefore an excellent tool to identify star-forming structures over a wide range of scales in complex, cluster-forming environments. There are multiple methods to decompose both 2-dimensional (position-position; PP) and 3-dimensional (position-position-velocity; PPV) datasets into coherent structures at various scales, including Gaussian deconvolution techniques, clump-finding algorithms, filament identification, and hierarchical algorithms, called dendrograms, that create a tree of structures by linking nested objects, and therefore capture the hierarchical nature of cloud structure. Here, we perform the latter analysis on the Serpens South \amm\ (1,1) data cube using the \emph{astrodendro} package\footnote{\url{http://www.dendrograms.org}}.  A detailed discussion of the methodology is presented by \citet{rosolowsky08d}. With this method, local emission maxima are identified in PPV space in the \amm\ datacube, and followed until they merge with each other, whereupon the merged structure is similarly tracked until it merges with another structure, repeating until the lowest structures' line brightness is less than some defined cutoff value. Whether or not structures are deemed to merge depends on several parameters, including a minimum difference in brightness required to allow two structures to remain separate that is generally a few times the rms noise of the dataset. The result is a structure tree, where one or more `roots', defining the structures with lowest line brightness, are connected to `branches' and eventually `leaves', the top-level structures. 

The \amm\ (1,1) line is composed of 18 separate hyperfine components that blend together into a single main and four satellite emission peaks. Running the dendrogram analysis on the original datacube would therefore add spurious structure to the results, as well as give inaccurate measures of, i.e., the velocity dispersion in the detected structures. Instead, we first created a simulated datacube from the fitting routine described in \S\ \ref{sec:line_fit} by determining, at each pixel, the line profile for a single hyperfine component given the best-fit results to the full (1,1) and (2,2) emission lines. We scale the single Gaussian component to the peak of the main \amm\ (1,1) component, and add back in the residual from the full hyperfine fit. There is some error introduced through this method, particularly toward regions where the \amm\ emission is composed of two separate, but closely spaced velocity components. Recall that our hyperfine fitting routine allows a single component only. If the two components are similar in brightness, the best-fit \vlsr\ tends to be intermediate to the individual \vlsr\ of the two components, while the line width $\Delta v$ will be broad. Where this behavior occurs, the residuals to the fit can be negative, resulting in spurious features in the deconvolved, residual-added cube. In Serpens South, several areas are affected at the edges of filaments and cores, visible in Figure \ref{fig:linefits} as narrow regions with sharp transitions in \vlsr\ and rapid increases in $\Delta v$. Apart from these regions, there is little evidence for the presence of multiple velocity components in the \amm\ data, but we note that for small velocity variations, the hyperfine structure of the \amm\ lines could conceal underlying velocity structure. To be sure, we checked carefully the dendrogram analysis results, and found that the threshold and interval values used were sufficient to avoid creating artificial features in these regions. 

\begin{figure}
\includegraphics[trim=0 50 0 0, clip, width=0.45\textwidth]{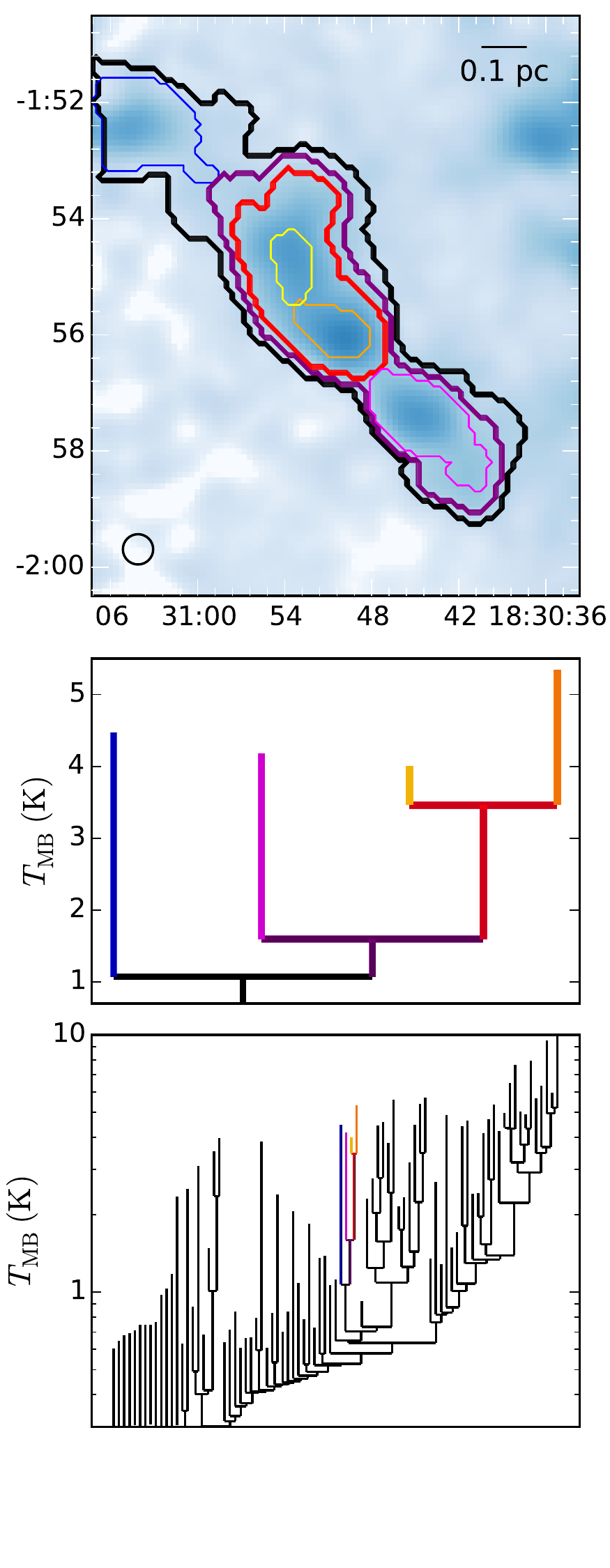}
\caption{An example of the results of the dendrogram analysis for a section of Serpens South. Top: the integrated \amm\ (1,1) intensity for a structure in the north-east of the map in Figure \ref{fig:mm11}. The contours highlight individual structures identified in the dendrogram analysis, projected from the 3D cube onto the 2D moment map as described in \S \ \ref{sec:masses}, causing some contours to overlap. The beam size and image scale are shown at the bottom left and top right, respectively. The edge of the map introduces a truncation of contours where the \amm\ emission clearly extends beyond the observed region. \referee{Middle: the structure tree for the highlighted structures only. Colors on the tree match the colors of the structures at top. Bottom: the full dendrogram tree for Serpens South. Colors on the tree match the colors of the structures at top.} \label{fig:dendro}}
\end{figure}

In our analysis, we limit the dendrogram routine to emission brighter than 0.3~K, or 5 times the mean rms noise, and mask the regions at the edge of the map with rms noise greater than 0.1~K. \referee{We require the minimum separation in brightness between the peak flux of  a structure and the value at which it is being merged into the tree to be 0.12\ K, or 2 $\times$ the rms noise. }We furthermore require that independent structures be at least as large as the 32\arcsec\, beam in projection over \referee{at least} two velocity channels (each 0.15~\kms), and have a peak line brightness of at least $0.6$\ K. The peak line brightness requirement removed several low-level structures primarily near the noisy edges of the map. Otherwise, the noise level in the map is relatively constant as noted in \S\ \ref{sec:obs}. Increasing the number of pixels required, or the minimum separation in brightness between non-merging structures, reduces the total number of structures identified as expected, but does not change significantly general trends in the data discussed further below. 

With these parameters, we obtain a dendrogram for Serpens South that contains a total of 155 identified structures, of which 85 are leaves, or top-level structures\referee{, and 15 are `roots'. In Figure \ref{fig:dendro}, we show an example of how part of the structure tree maps onto the \amm\ (1,1) integrated intensity. In the top panel, contours highlight individual dendrogram structures projected onto the 2D \amm\ (1,1) integrated intensity map, where the contour colors are matched to the corresponding `branch' and `leaf' structures in a zoomed-in portion of the full structure tree (middle panel). The structure tree does not show any spatial information about the structures, but instead identifies the peak structure line brightness temperature (the top of each vertical line), and the line brightness temperature at which each structure merges with adjacent structures (horizontal lines joining vertical lines). We show the full structure tree for Serpens South in the bottom panel of Figure \ref{fig:dendro}.} Most of the structures in Serpens South merge to a single root, corresponding to the large filamentary structure connected to the central YSO cluster, but a significant fraction remains separate in PPV space to our brightness temperature lower limit. In some cases, structures may be artificially disconnected as a result of our limited map extent, since at low brightness temperatures, emission frequently extends to the edges of the map. In other cases, there are clear separations in structure velocities that preclude the connection of all structures, as can be seen in the $v_\mathrm{LSR}$ \amm\ line fitting results shown in Figure \ref{fig:linefits}. 

For each structure, we obtain multiple parameters. Of most interest here are the mean positions of the structures in $x$, $y$, and $v$, the major and minor axes of the structure's projection onto the PP plane, computed from the intensity weighted second moment in the direction of greatest elongation in the PP plane ($\sigma_\mathrm{maj}$ and $\sigma_\mathrm{min}$), and the position angle ($\theta_\mathrm{PA}$). The effective radius of each structure is defined as the geometric mean of the major and minor axes, $R_\mathrm{eff} = (\sigma_\mathrm{maj} \ \sigma_\mathrm{min})^{1/2}$. 

Some regions, particularly the north-south main filament, are less fragmented than others, such as the filamentary structures in the north-east. This difference may be partly due to both the 32\arcsec\ resolution of the GBT data and the broader line widths seen in the  region, which could mask fragmented structures separated by small shifts in $v_\mathrm{LSR}$. In the southern filament, however, substructure is clearly visible in extinction in Figure \ref{fig:mm11}. Similar features are not apparent in the north, evidence that the lower fragmentation seen in the dendrogram analysis toward this structure is real. The position angles of the leaves are largely correlated with the direction of elongation of the larger-scale structures in which they are embedded. 

\subsubsection{Structure masses and densities from continuum emission}
\label{sec:masses}

In this section, we determine the masses of the structures identified in the dendrogram analysis. Previous calculations of structure masses from dendrogram analyses of CO datacubes have used a conversion factor between the abundance of CO and H$_2$ to determine masses directly from the total CO flux in each structure \citep{rosolowsky08d,goodman09,pineda09}. We showed above that the \amm\ abundance varies across the dense gas in Serpens South. Furthermore, variations in the line excitation and kinetic temperature across the region disallow the direct conversion from flux to mass, even assuming a constant \amm\ abundance relative to H$_2$.  

\referee{We thus determine instead the masses and densities of the dendrogram-identified structures from the Herschel \nh\ map of the Serpens Aquila region \citep{konyves15}, rather than directly from the \amm\ data. The 36\arcsec\, resolution (FWHM) of the \nh\ map is well-matched to the GBT beam. We first re-gridded the \nh\ data to match the \amm\ moment map. We then project the 3D dendrogram structure to the position-position plane to create a 2D mask for each structure, and determine the mass in the resulting mask from the \nh\ data assuming $d = 429$~pc. The total mass for each structure thus includes all material along the line of sight. As stated earlier, there are only a few regions where there is clear overlap of structures in velocity space along the same line-of-sight, and we thus expect the projection of structure from 3D to 2D will introduce little error in the mass calculations. The masses of branch and root structures also include the masses of leaves and branches within them. }

Many structure-identifying routines, such as Gaussian deconvolution routines, identify cores within clouds and along filaments, where large-scale background structure is first subtracted from the image before core sizes and masses are determined. Since we have identified structures in PPV space, it is not clear how best to map the `background' level derived from the \amm\ cube to a `background' level in the continuum. \referee{The masses of structures embedded within larger features, such as cores within filaments, may therefore be smaller than indicated by this analysis. We discuss this further in \S\ \ref{sec:cont}, where we compare the \amm\ structures with cores identified in continuum emission where large-scale features have been removed. }

\begin{figure}
\includegraphics[trim=5 0 3 5,clip,width=0.48\textwidth]{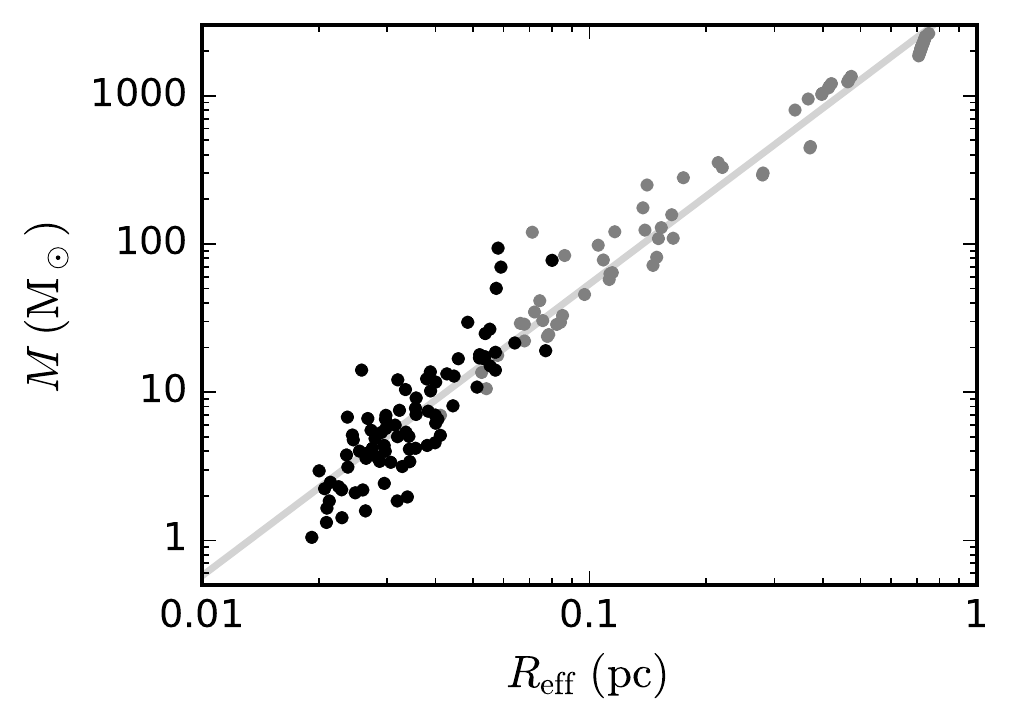}
\caption{The mass versus radius relationship for structures in Serpens South. Black points indicate the top-level structures in the dendrogram analysis, while grey points show the dendrogram branches and roots. The grey line shows the best fit to the data, where we find a power law index of $1.99 \pm 0.03$. \label{fig:m_vs_r}}
\end{figure}

We show in Figure \ref{fig:m_vs_r} the resulting mass-radius relation for the dendrogram structures in Serpens South, with the top-level leaves highlighted in black. We find \referee{that $M \propto R^2$, the relation typically measured in molecular clouds \citep[e.g.,][]{larson81}, where a slope of $1.99 \pm 0.03$ best fits the data, with an r-value of 0.96. }Shown by the grey line fit to the data, the relation holds over \referee{approximately} three orders of magnitude in mass ($\sim 1 \ \mathrm{M}_\odot \lesssim M_\mathrm{struc} \lesssim 1000 \ \mathrm{M}_\odot$) and a factor of 50 in radius ($\sim 0.02 \ \mathrm{pc} \lesssim R_\mathrm{eff} \lesssim 1 \ \mathrm{pc}$). While this result is similar to that found by \citet{larson81}, suggesting \nh\, is constant, comparison between \amm\ emission and $N(\mbox{H}_2)$ shows that \amm\ emission, to our sensitivity limits, is present over a factor of ten in H$_2$ column density. The similarity to Larson's relation may be instead an artifact of the \amm\ emission preferentially highlighting structures well above the peak of the cloud column density probability distribution function \citep{ballesteros12}. \referee{Alternatively, if the masses of the smallest structures are overestimated as discussed above, the relationship may be steeper than indicated here; however, the mass calculations of the grey points (which agree well with the $M \propto R^2$ line) should be unaffected. We note that some of the data points at large $R_\mathrm{eff}$ in Figure \ref{fig:m_vs_r} and following plots represent merging of small features with larger-scale structure, and are not entirely independent. As a result, the overall trend may be less significant than is suggested by the Figure. In general, the lack of independence of these few points does not impact the results that will presented in the following discussion. }

\section{Discussion}
\label{sec:discussion}

\subsection{Structure of the SSC and connected filaments}
\label{sec:struc}

\begin{figure}
\includegraphics[trim=0 0 5 5, clip, width=0.47\textwidth]{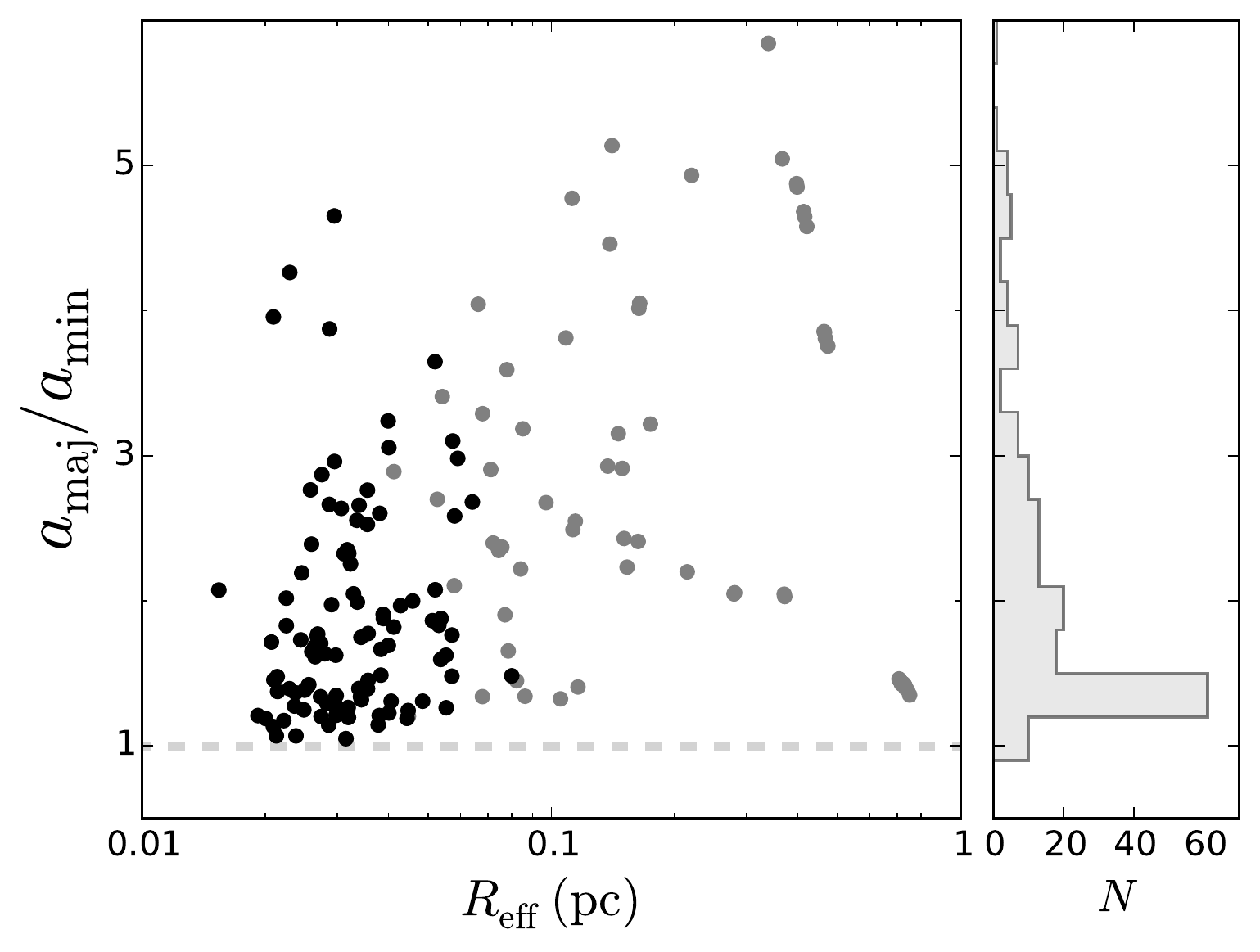}
\caption{Aspect ratio as a function of effective radius, $R_\mathrm{eff}$, for all structures identified in the dendrogram analysis in Serpens South. Black points indicate the top-level structures in the dendrogram analysis, while grey points show the dendrogram branches and roots. To the right, a histogram of the aspect ratios shows that most structures have an aspect ratio less than two, and primarily this is true for structures with $R_\mathrm{eff} \lesssim 0.1$~pc.  \label{fig:aspect}}
\end{figure}

Although dendrogram analysis is not designed to detect filamentary structure, we can identify coherent, elongated structures through examination of the structures' aspect ratios. We show in Figure \ref{fig:aspect} the aspect ratio of the dendrogram-identified structures as a function of effective radius, $R_\mathrm{eff}$. On average, there is an increase in the aspect ratio with increasing structure size, such that larger structures tend to be more elongated, while most structures with $R_\mathrm{eff} \lesssim 0.1$ \ pc have aspect ratios between one and two, typical for star-forming cores.  Above an effective radius $R_\mathrm{eff} \sim 0.1$~pc, nearly all structures have aspect ratios greater than two, and in some cases are substantially elongated with $\sigma_\mathrm{maj} \ / \ \sigma_\mathrm{min} \sim 4-6$. \referee{As noted previously, some data points represent the merging of small features with large-scale structures and are not independent, but these can be clearly identified in the Figure as points that cluster along a compact trend line (such as the grouping of points around $R_\mathrm{eff} \sim 0.4$\ pc with aspect ratios decreasing from $\sim 5$, for example).} On the largest scales, the structures include much of the \amm\ emission in Serpens South and thus return to a small aspect ratio. This effect is due to a combination of \referee{the approximately square map area} and the overall velocity coherence of much of the star-forming region at low line brightness.

\edits{Within the Serpens South filaments, we do not see evidence of the bundled, velocity-coherent filaments 0.5\ pc in length identified by \citet{hacar13} toward Taurus L1495/B213. This difference may be a resolution effect, both spatially and kinematically, since separate filamentary structures subsonically separated in velocity space would be indistinguishable in our data due to the many hyperfine components of the \amm\ inversion transitions. The velocity separations of $\sim 1 - 1.5$\ \kms\ observed between bundled filaments in Taurus, however, would have been easily resolved in our data, if present. }

\subsubsection{Continuum structure comparison and protostellar population}
\label{sec:cont}

Through the dendrogram analysis, we identify 155 structures in \amm\ emission within Serpens South, where 85 of these are top-level structures, or leaves. We next compare the results of the 3D dendrogram analysis with the Aquila dense core catalog from the Herschel Gould Belt Survey \citep{konyves15}. We identify matched \amm\ leaves and Herschel cores when the Herschel core centre is within $1 \ R_\mathrm{eff}$ of the \amm\ leaf centre. Of the 85 \amm\ leaves identified in Serpens South, we find 75 that correspond with Aquila cores. Although the continuum analysis does not have any kinematic information, \citeauthor{konyves15} find that $60\% \pm 10\%$ of starless dense cores in Aquila are likely gravitationally bound and prestellar, suggesting they will eventually form a star or group of stars, by comparing the observed core mass with the critical Bonnor-Ebert mass. Of the 75 \amm\ leaves that overlap with continuum-identified cores, 10 are identified by \citeauthor{konyves15} as protostellar based on 70~\micron\ emission, and 56 of the remaining 65 leaves are classified as prestellar. The \amm\ leaves thus highlight a significantly greater fraction of prestellar cores (86\%) than the continuum observations alone. This value is similar to the fraction of prestellar cores that are closely associated with filamentary structures in the continuum data ($75\%^{+15\%}_{-5\%}$). While we find low line brightness \amm\ emission over most of Serpens South, the top-level structures identified in the dendrogram analysis are located primarily along the continuum-identified filaments (see Figures 3 and 4 of  \citeauthor{konyves15}). The \amm\ emission therefore is predominantly associated with the denser filamentary structures, where prestellar cores are more likely to be found. 

While many of the \amm\ leaves are correlated with continuum cores, the Herschel continuum cores have smaller reported physical sizes \referee{and masses}. When corrected for the 429\ pc distance used for Serpens South in this paper, however, we find that the range of core sizes agree well with the effective radii of the dendrogram leaves, although the direct core-to-leaf comparison has substantial scatter. This lack of one-to-one correlation is not unexpected, given our analysis uses both a different structure identification technique and a different dense gas tracer than \citeauthor{konyves15}. The mass-size relationship shown in Figure \ref{fig:m_vs_r} thus overlaps that found in the continuum analysis for small structures, and furthermore extends it to larger sizes and masses for the underlying large-scale structures identified in the \amm\ dendrogram. The substantial overlap between \amm\ and continuum structures emphasizes three things: (i) the continuum analysis indeed highlights high density structures in Serpens South, rather than high column density; (ii) \amm\ is an excellent tracer of high density continuum structures; and (iii) as noted above, structures that overlap along the line-of-sight, but with different velocities, are not contributing substantially to confusion or omissions in structure catalogues at this resolution in Serpens South.  

\subsubsection{The line width-size relation in dense gas}
\label{sec:sigma_r}

The dendrogram analysis itself provides measures of $R_\mathrm{eff}$ and $\sigma_v$ for each structure, by calculating the intensity-weighted second moment of velocity within the 3D mask determined for each leaf, branch and root. Different results can be obtained, however, depending on how the structure properties are measured \citep[for more details, see ][]{rosolowsky08d}. In particular, the velocity dispersion for the `leaves' of the dendrogram, which are identified based on the brightest portion of the \amm\ line, can be significantly underestimated due to including only the brightest portion of the spectral line in the calculation. To mitigate this, we instead determine the velocity dispersion of each structure from the \amm\ fitting, following the same method as for the mass calculation in \S\ \ref{sec:masses}: \referee{we define a 2D mask based on the projection of the 3D dendrogram structure to the position-position plane, and calculate the mean velocity dispersion of all pixels within the mask from the \amm\ $\Delta v$ fit results, weighted by the integrated intensity of \amm\ (1,1) at each pixel.} We then use the mean gas temperature for each structure to determine the non-thermal velocity dispersion, $\sigma_\mathrm{nt} = \sqrt{\sigma_v^2 - k_\mathrm{B} T_\mathrm{K} / m_{\mathrm{NH}_3}}$. Where robust gas temperature measurements were not found, we set $T_\mathrm{K} = 11$\ K, the mean value across Serpens South. This assumption is likely reasonable, as we find no trend with \tk\ and size scale.

\begin{figure}[t]
\includegraphics[trim=5 0 4 0,clip,width=0.47\textwidth]{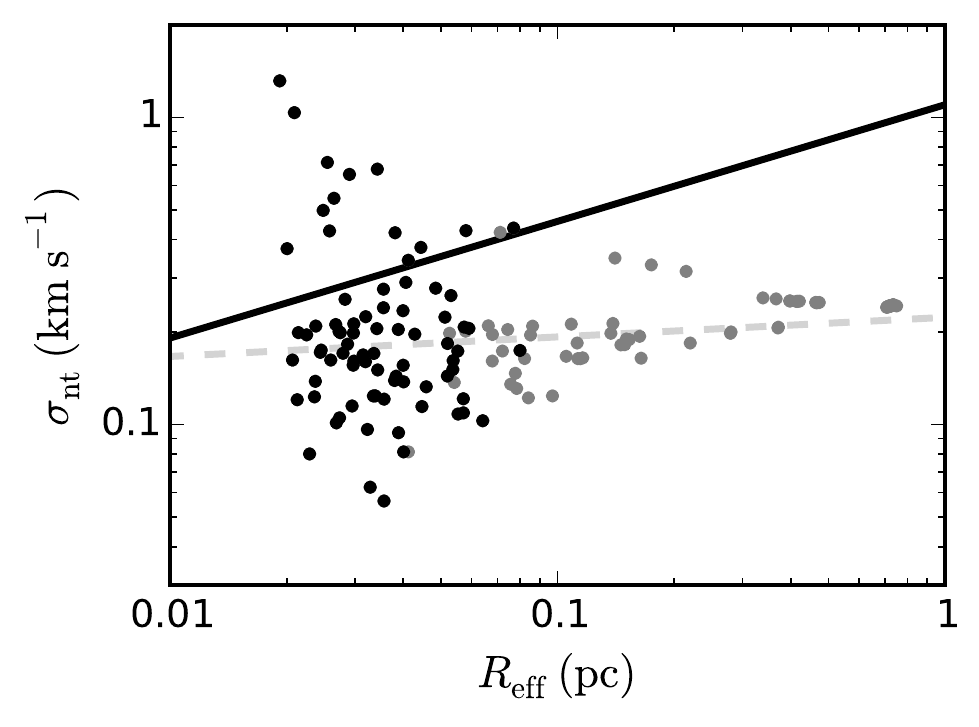}
\caption{The line width-size relation for structures in Serpens South. Black points indicate the top-level structures in the dendrogram analysis, while grey points show the dendrogram branches and roots. The black line shows the relationship determined by \citet{larson81}. The dashed grey line shows the fit to the \amm\ data, finding a slope consistent with zero. \label{fig:dv_vs_r}}
\end{figure}

We show in Figure \ref{fig:dv_vs_r} the resulting size-line width relation. The power law index is consistent with zero ($0.03 \pm 0.03$; dashed grey line), with a typical value $\sigma_\mathrm{nt} = 0.22$\ \kms. This result is significantly different from the canonical power-law indices of $\sim 0.4 - 0.5$ found in molecular clouds using molecular tracers of lower density \citep[e.g.,][]{larson81,solomon87}. \referee{The original Larson relation between size and line width was determined with relatively poor angular resolution, such that the observed trend in line width with structure size was due to both a real rise in the gas velocity dispersion as well as increased dispersion in the line-of-sight velocity of the gas. To the limit of our angular resolution, we have removed the line-of-sight component in our analysis by using the \amm\ fit results. Similar analyses of dense gas (traced by \dia\ at 7\arcsec\ resolution) in the Serpens Main and Barnard 1 clouds have also found shallow trends in line width with increasing structure size, and have shown that the variation in \vlsr\ increases with size scale \citep{klee14,storm14}. Given the $\sim 2$\ \kms\ variation in \vlsr\ present across Serpens South, observations with poor angular resolution would result in a greater slope in the size-line width relation. }

Flat size-line width relations found for higher density tracers may reveal the \referee{threshold spatial scale} at which \referee{the turbulent motions characteristic of the lower-density gas dissipate} \citep{goodman98}. Most of the dense gas traced by \amm\ is characterised by \referee{small (sub- or transsonic) non-thermal motions:} at the mean gas temperature in Serpens South of 11\ K, the sound speed $c_s =  0.20$\ \kms, comparable to the mean non-thermal velocity dispersion over the entire region, $\sigma_\mathrm{NT} = 0.22$\ \kms. \citeauthor{goodman98} find, however, that this transition to coherence occurs on size scales of $\sim 0.1$\ pc, whereas in Serpens South we find subsonic $\sigma_\mathrm{NT}$ values over regions of length $\sim 1$\ pc. The \citeauthor{goodman98} result is derived from more or less isolated dense cores, and the specific size scale at which a transition to coherent motions may have been dictated by their target sample. Furthermore, Serpens South is a filamentary region on the cusp of a burst of low mass star formation. The much larger-scale coherent structure found here could be due both to its formation history and its relative youth. Some simulations of filamentary structures formed from supersonic shocks within turbulent molecular clouds predict that molecular line tracers of high density gas show narrow line widths and small line-of-sight velocity variations over the length of the filament, in agreement with our observations \citep{rsmith12}. This behavior is due to the dissipation of turbulence in the dense, post-shock gas. Figure \ref{fig:linefits} further shows that regions \referee{with subsonic non-thermal motions} tend to be starless, and are thus not yet impacted significantly by protostellar outflows that might drive turbulence in the gas. 

 \referee{\citet{storm14} argue that shallow trends in the velocity dispersion of dense gas (traced by \dia\ or \amm, here), compared with an increase in the dispersion of \vlsr\ values with size, indicates that the depth of the dense gas along the line of sight is similar across the cloud. The lack of correlation between structure size and velocity dispersion in Serpens South may then suggest that the filamentary structures, like the main north-south filament extending from the central cluster, are not substantially inclined along the line-of-sight.}

\subsection{Stability of Cores and Filaments}
\label{sec:stability}

\subsubsection{Virial analysis of \amm\ hierarchical structure}
\label{sec:virial}

We next investigate the stability of the structures traced by \amm\ emission. In the absence of external pressure or magnetic fields, the virial mass, \mvir, for each structure is a measure of how much mass can be supported against self-gravity by the combination of thermal and non-thermal motions in the gas. We then define \mvir\ as

\begin{equation}
M_\mathrm{vir} = \frac{5 \sigma^2 R}{a G},
\end{equation}

\noindent where $G$ is the gravitational constant, and

\begin{equation}
a = \frac{1-p/3}{1-2p/5}
\end{equation}

\noindent accounts for a power-law radial density profile of the structure \citep{bertoldi92}, where we set $p = 1.5$. For comparison, the density profile of a Bonnor-Ebert sphere is flat to some radius, then declines as $r^2$. We set $R = R_\mathrm{eff}$.

The velocity dispersion $\sigma$ includes both the thermal and non-thermal motions of the particle of mean mass, where $<m> = \mu m_\mathrm{H}$ and $\mu = 2.33$ for molecular gas. We calculate $\sigma$ from the \amm\ fit results following

\begin{equation}
\sigma^2 = \sigma_v^2 -  \frac{k_\mathrm{B} T}{m_\mathrm{NH3}} + \frac{k_\mathrm{B} T}{\mu m_\mathrm{H}}, 
\label{eqn:sig}
\end{equation}

\noindent \referee{where $\sigma_v$ is the velocity dispersion calculated for each structure in \S\ \ref{sec:sigma_r},} $m_\mathrm{NH3}$ is the mass of \amm, $m_\mathrm{H}$ is the mass of a hydrogen atom, and $k_\mathrm{B}$ is Boltzmann's constant. As before, we set $T$ to the \amm-derived gas temperatures $T_\mathrm{K}$ for each structure. The virial parameter, $\alpha_\mathrm{vir}$, is then the ratio of the virial mass to the structure mass, $\alpha = M_\mathrm{vir}/M$. Below a critical value $\alpha_\mathrm{vir} \lesssim 2$, structures are not able to support themselves against self-gravity, in the absence of magnetic pressure, and are unstable to collapse. 

\begin{figure}
\includegraphics[trim=0 0 0 0,clip,width=0.48\textwidth]{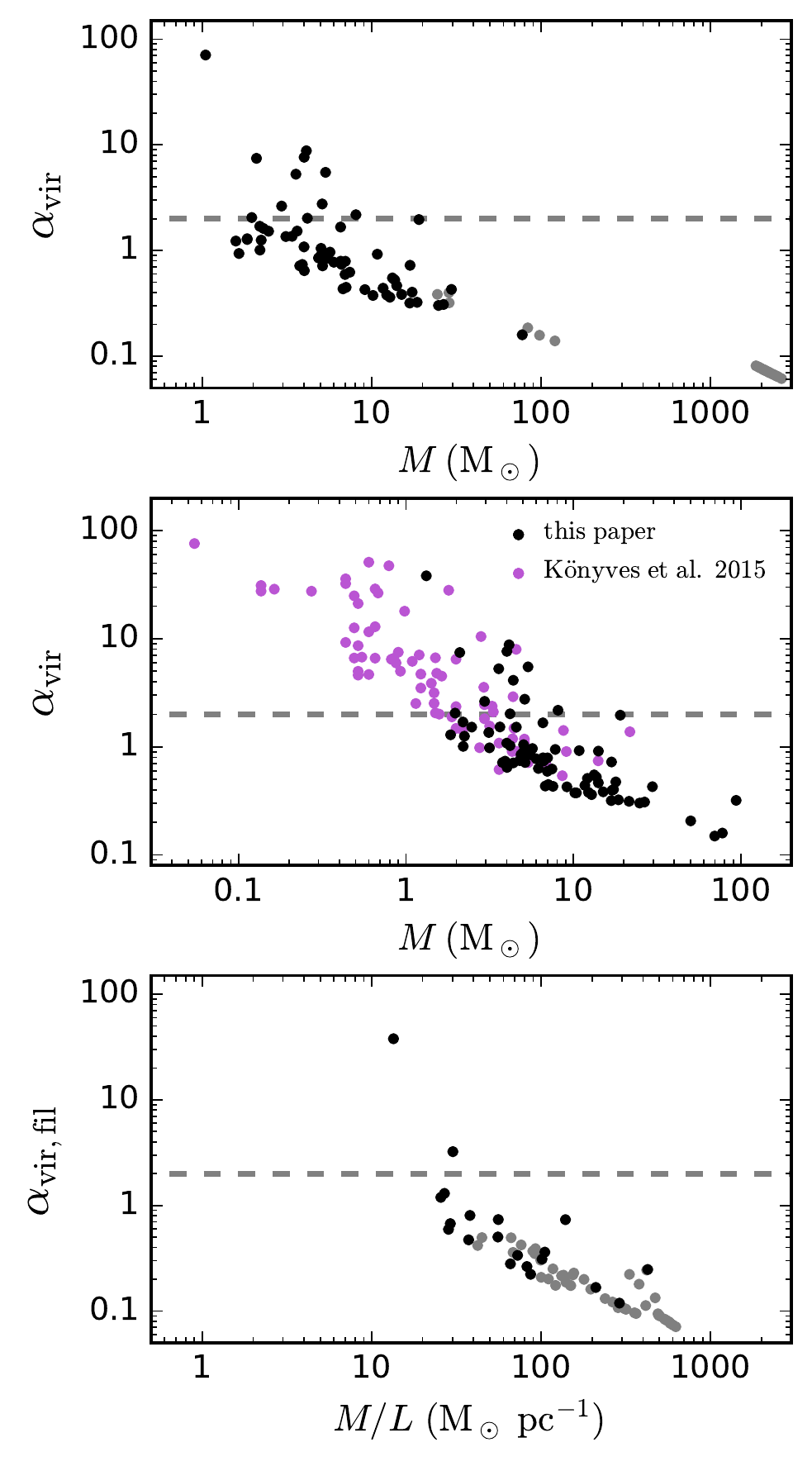}
\caption{Top: Virial parameter as a function of structure mass for structures with aspect ratios less than two. In the top and bottom panels, black points identify top-level dendrogram structures (`leaves') while grey points identify branches and roots in the dendrogram. \referee{Middle: Where structures overlap, a comparison of the virial parameter of dendrogram leaves for masses determined using the \amm\ 2D masks on the continuum-derived $N(\mathrm{H}_2)$ maps (black points), and those from \citet{konyves15}, where large-scale emission is removed from core fluxes before calculating cores masses (purple points). }Bottom: Filamentary virial parameter as a function of structure mass per unit length for structures with aspect ratios $\geq 2$.  
\label{fig:m_vs_alphaVir}}
\end{figure}

The virial analysis assumes spherical symmetry. Consequently, in Figure \ref{fig:m_vs_alphaVir} (top), we show $\alpha_\mathrm{vir}$ as a function of structure mass in Serpens South for structures with aspect ratios of two or less. We find a strong trend of decreasing virial parameter with increasing structure mass. \referee{Most of the structures in Serpens South lie below the critical $\alpha_\mathrm{vir}$ value, with only a scattering of the top level structures with $\alpha_\mathrm{vir} > 2$. Of the protostellar structures in Serpens South, all but one have virial ratios below the critical value.} On the largest scales, the most massive structures (essentially the entire hub and filament structure of the region) are characterized by extremely low virial parameters, $\alpha_\mathrm{vir} < 0.1$. In a review of virial analyses toward multiple molecular clouds, \citet{kauffmann13} find the same result for high-mass star-forming regions. 

\referee{As noted in \S\ \ref{sec:masses}, the masses calculated in this paper, based on a 2D mask of the individual 3D dendrogram structure, include all material along the line of sight, and do not subtract any underlying large-scale emission. This is in contrast with the analysis done by \citet{konyves15}, where submillimeter cores are identified after filtering the continuum maps. As noted previously, these filtered masses are generally lower than those derived in this work. We show, then, in the middle panel of Figure \ref{fig:m_vs_alphaVir}, a comparison between the virial parameter calculated above and the virial parameter using the submillimeter-derived mass as a function of structure mass for the top-level dendrogram leaves that correlate with submillimeter cores in the \citeauthor{konyves15} catalogue. While the results overlap, the lower masses stemming from the filtered submillimeter analysis shift the virial parameters of a substantial fraction of the structures above the critical value, such that many are consistent with being unbound in the absence of external pressures. }

Significant external pressure contributions include the weight of the larger molecular cloud on the embedded structures, as well as ram pressures from cloud collapse and gas accretion. In the Taurus B211/213 filaments, for example, \citet{seo15} show that the internal pressure of \amm\ dendrogram-identified leaves is very similar in magnitude to the estimated surface pressure due to the weight of the filaments on the leaves ($8 \times 10^5$ \ K cm$^{-3}$ and $6 \times 10^5$\ K cm$^{-3}$, respectively). Furthermore, the authors estimate that the ram pressure due to filamentary accretion \citep[suggested by velocity gradients in CO data;][]{palmeirim13} is within a factor of two of the internal leaf pressure, and conclude that most of the structures that are unbound based on a pressure-free virial analysis are indeed bound by external pressures. Similarly, \citet{heitsch13} show that ram pressures predicted by a cloud collapse model match observations and are sufficient to confine cores in the Pipe nebula. 

We can estimate the effect of these pressure terms on structure stability in Serpens South. For infalling gas, the ram pressure $P_\mathrm{ram} = \rho v_\mathrm{in}$. In \papi, we fit asymmetric self-absorbed HCN line profiles toward the northern filament in Serpens South with a simple analytic infall model, finding radial infall speeds of $\sim 0.5$\ \kms, while \citet{kirk13} find similar infall speeds toward the filament south of the central cluster. Assuming a density lower limit of $n \sim 10^4$\ \cc\ (i.e., the lower bound where \amm\ (1,1) and (2,2) are typically excited), ram pressures $P/k_\mathrm{B} \sim 4 \times 10^5$\ \cc\ K are expected. Densities at the filament are likely greater, since they are visible in higher density gas tracers such as \dia\ (\citeauthor{kirk13}), so ram pressures in Serpens South due to accretion may be on the order $P/k_\mathrm{B} \sim 10^6$\ \cc. 

We next estimate the internal pressure of the Serpens South leaves, where $P_\mathrm{int} = M \sigma^2 / V$, \referee{using the Herschel-reported masses, and} assuming the structures are prolate spheroids with a volume $V = \frac{4}{3} \pi \sigma_\mathrm{min}^2 \sigma_\mathrm{maj}$. We find a median $P_\mathrm{int}/k_\mathrm{B} = 1.2 \times 10^7$\ K cm$^{-3}$, an order of magnitude greater than the estimated ram pressure due to accretion. For a density structure similar to a critical Bonnor-Ebert sphere, however, the surface pressure is $\sim 40$\ \% of the mean internal pressure, giving a median value at the leaf surface of $P/k_\mathrm{B} \sim 5 \times 10^6$\ K cm$^{-3}$. We do not directly measure the effect of the weight of the molecular cloud, but in clouds like Orion B and $\rho$ Ophiuchus that are better analogues of Serpens South than Taurus, surface pressures of cores are estimated to be $P/k_\mathrm{B} \sim 10^6 - 10^8$\ K cm$^{-3}$ \citep{johnstone00, johnstone01}, in agreement with our estimate of the internal leaf pressures. Given these values, it is clear that \referee{even using the Herschel-reported masses,} many of the structures that appear unbound in Figure \ref{fig:m_vs_alphaVir} are likely pressure-confined objects. The relative importance of the weight of the surrounding cloud versus pressure from accretion requires a more detailed analysis than we can do with the data presented here.

\subsubsection{Stability of elongated (filamentary) structures}

We noted in \S\  \ref{sec:struc}, and showed in Figure \ref{fig:aspect}, that many of the structures in the size range between $0.1 \lesssim R_\mathrm{eff} \lesssim 0.6$~pc are significantly elongated. The above virial analysis is calculated for spherically symmetric objects, and does not provide an accurate measure of the stability of filamentary structures. For elongated structures, we instead compare the mass per unit length with the virial mass per unit length, in the absence of magnetic support, defined as \citep{fiege00}

\begin{equation}
m_\mathrm{vir, fil} = \frac{2 \sigma^2}{G}, 
\end{equation}

\noindent where we determine the velocity dispersion $\sigma$ from the \amm\ line fitting as in Equation \ref{eqn:sig}. For each structure, we calculate $M/L$ using the mass determined above, and setting $L = 2\sqrt{2 \mathrm{ln} 2} \ \sigma_\mathrm{maj}$, or twice the FWHM of the major axis of the structure from the dendrogram analysis. In the bottom panel of Figure \ref{fig:m_vs_alphaVir}, we show $m_\mathrm{vir,fil}$ as a function of $M/L$ for all structures with aspect ratios $\geq 2$. 

Figure \ref{fig:m_vs_alphaVir} shows that most elongated structures are supercritical under the filamentary virial analysis, in the absence of additional support, with a minimum $\alpha_\mathrm{vir,fil} \sim 0.06$ on the largest scales. In the filamentary case, supercritical filaments are unstable to radial collapse. 

Several structures have virial parameters significantly above the critical value, and these correspond to the highly elongated leaves visible as outliers at small radii in Figure \ref{fig:aspect}. In the \amm\ maps, these structures tend to be located away from the larger-scale filamentary structures in Serpens South. 

\subsubsection{Impact of magnetic fields}

The low virial parameters discussed above imply that on large scales, dense gas structures in Serpens South are greatly unstable to gravitational collapse, unless there are substantial forces acting against gravity, of which magnetic fields are the prime candidate. For magnetized clouds of radius $R$ with a magnetic flux $\Phi$, the critical mass is then $M \sim M_{B=0} + M_\Phi$, where \citep{tomisaka88}

\begin{equation}
M_\Phi = 0.12 \frac{\Phi}{G^{1/2}} = 0.12 \frac{\pi <B>R^2}{G^{1/2}}  . 
\end{equation}

Here, $<B>$ is the mean magnetic field strength. \citet{kauffmann13} show that in the presence of a magnetic field, the critical virial parameter can be rewritten as

\begin{equation}
\alpha_\mathrm{crit, B} = \frac{2}{1 + M_\Phi/M_\mathrm{vir}}, 
\end{equation}

\noindent such that with sufficiently high magnetic field strengths, the critical virial parameter can become quite low. As mentioned previously, \citet{sugitani11} show that the magnetic field is relatively ordered in the lower density material surrounding the dense filaments, with magnetic field lines approximately perpendicular to the main filament running north and south of the central cluster. Assuming a distance of 260~pc toward the region, and that the magnetic field is almost perpendicular to the line of sight, the authors estimate the magnetic field strengths in the plane of the sky, $B_\mathrm{pos} \sim 150-200 \ \mu$G using the Chandrasekhar-Fermi method. Correcting for the greater distance used in this paper, where $B_\mathrm{pos} \propto d^{-1/2}$, we revise this measurement to $<B_\mathrm{pos} > \sim 120-150 \ \mu$G. On the largest scales, we find structures with $M_\mathrm{vir} \sim 170$~M$_\odot$ and radius $R_\mathrm{eff} = 0.75$~pc. The corresponding virial parameter is $\sim 0.06$. Solving for $M_\Phi$ with $<B> = 150 \ \mu$G, we find the critical virial parameter to be $\alpha_\mathrm{crit,B} = 0.23$. This value is still a factor of $\sim 4$ greater than needed for the structure on largest scales to be stable. We note that \citet{tanaka13} also adjust the $B_\mathrm{los}$ estimate in Serpens South to account for an inclination of 45\arcdeg\ between the magnetic field and the plane of the sky, with a resulting overall magnetic field strength of only $B \sim 80 \ \mu$G. This is substantially lower than required to provide stability to the filament. While it thus appears that the magnetic field strength is insufficient to support the filaments against collapse, no measurements have yet been made of the magnetic field strength or direction in the denser filamentary gas. Such data are needed to evaluate fully the influence of the magnetic field on the stability of the star-forming region. 

\subsubsection{Implications of instability}

If the extremely low virial parameters in Serpens South imply that the region is globally collapsing, then why is the \amm\ velocity dispersion transsonic, on average, over large spatial scales? During spherical free-fall collapse, observed line widths increase due to turbulent motions driven by collapse, and virial parameters should remain close to unity \citep{larson81}. Furthermore, observed radial infall motions of $\sim 0.5$\ \kms\ observed toward the northern and southern filaments are supersonic \citep[\papi; ][]{kirk13}, and close to the 0.6\ \kms\ infall speed predicted for an isothermal cylinder undergoing freefall collapse \citep{heitsch13}. \citet{matzner15} point out, however, that where collapse is dominated by filaments, flows are identified by velocity gradients, and $\alpha_\mathrm{vir}$ will remain low when derived using the mean velocity dispersion of the fitted lines \citep[as was done here and by][]{kauffmann13}. Hydrodynamic simulations of filament formation within a turbulent medium, where filaments form from supersonic shocks, suggest that the higher density gas within filaments remains subsonic or transsonic \citep{gong15}. Following their hydrodynamic simulation with a radiative transfer analysis, \citet{rsmith12} showed that molecular tracers of high density gas toward collapsing cores in filamentary structures are emitted from the post-shock gas, with consequently sonic or subsonic velocity dispersions similar to those observed in Serpens South. Emission from molecular tracers of lower density gas \referee{should trace the more turbulent outer envelope of the filaments, and} show large line widths and variations in line-of-sight velocity. Additional studies of filamentary collapse models that include expectations for the excitation and kinematics shown by different molecular tracers are needed to explain fully both the large infall motions and small non-thermal motions seen toward the dense filaments in Serpens South. 

\begin{figure}[t]
\includegraphics[width=0.48\textwidth]{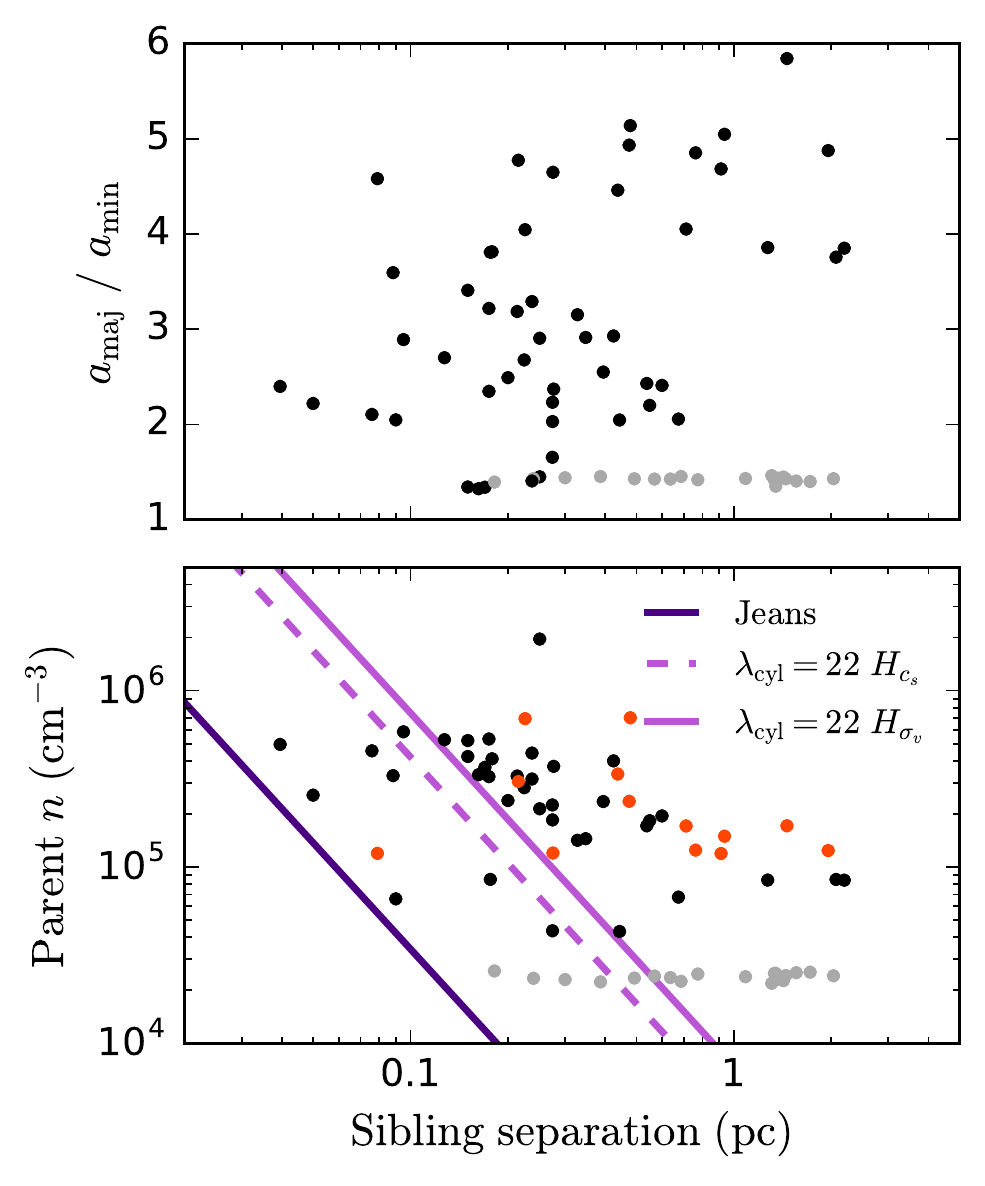}
\caption{Relationship between the distance between sibling structures (i.e., structures originating from the same parent structure, where pairs of structures merge at some line brightness to form a larger structure; see Figure \ref{fig:dendro}) and the aspect ratio (top) and the number density of the parent structures (bottom). At top, grey points indicate incidences where the mean density of the parent structure is just slightly greater than the mean density of the entire Serpens South region. These small structures merge at low line brightnesses with the main cloud, and are thus not tracing fragmentation of individual features within the region. At bottom, black and grey points are as above, while orange points highlight incidences where the parent structure is elongated, with an aspect ratio greater than 4. The solid purple line shows the expected separation of structures undergoing Jeans fragmentation. The lighter purple lines show the expected fragmentation for an isothermal cylinder assuming only thermal support (dashed) and turbulent support (solid), where the velocity dispersion for turbulent support is the mean \amm-derived value over the entire region. \label{fig:frag}}
\end{figure}

\subsection{Fragmentation}
\label{sec:frag}

\subsubsection{Fragmentation in equilibrium structures}

Molecular gas supported purely by thermal gas motions is unstable on scales greater than the Jeans length, $\lambda_\mathrm{J} = c_s (G \rho)^{-1/2}$ in the spherical case, where $c_s$ is the sound speed and $\rho$ is the mass density of the gas. Where thermal support dominates in clouds significantly larger than the Jeans length, we therefore expect to see cloud fragmentation, where the embedded molecular gas clumps are physically separated approximately by $\lambda_\mathrm{J}$. 

In the filamentary case, cylindrical structures are expected to fragment on characteristic length scales that are equal to the fastest-growing unstable mode of the fluid instability, and differ from predictions from the spherical Jeans case. For an infinite, isothermal cylinder, the maximum fragmentation length scale is $\lambda_\mathrm{cyl} = 22 H$, where $H$ is the scale height of the cylinder \citep{nagasawa87,inutsuka92}. Thermally-supported cylinders embedded within an external medium will have a fragmentation length scale that depends on the ratio between $H$ and $R$, the cylinder radius: where $R \gg H$, $\lambda_{cyl} \sim 22 H$, while for $H \gg R$, $\lambda_{cyl} \sim 11R$ \citep{jackson10}. For a purely thermally-supported cylinder, the scale height is given by \citep{ostriker64} 

\begin{equation}
H_{c_s} = \sqrt{\frac{c_s^2}{4\pi G \rho_c}}
\end{equation}

\noindent where $\rho_c$ is the central mass density of the filament. In the case where turbulent motions add support to the filament, the sound speed $c_s$ is replaced by the velocity dispersion $\sigma_v$ ($H_{\sigma_v}$). For several individual filaments, \citet{jackson10} and \citet{kainulainen13} showed that the observed fragmentation length scale is more consistent with the turbulent filamentary fragmentation described above, rather than the prediction from thermal Jeans analysis. Here, we investigate the fragmentation of dense gas over the entire Serpens South cloud. 

We use the dendrogram analysis to identify structures that have fragmented from the same parent structure (siblings) and examine the fragmentation of the dense gas within Serpens South. We compare our observed fragments with predictions from both the spherical Jeans as well as filamentary fragmentation analysis. \referee{We calculate the mean density $\rho = M/V$ of the parent structures using the continuum-derived masses $M$ and the volume $V$ calculated as in Section \ref{sec:virial}, assuming the structures are prolate spheroids. }The number density $n = \rho / (\mu m_\mathrm{H})$, where $\mu = 2.8$. 

We show in Figure \ref{fig:frag} (top) the projected distance between sibling structures versus the aspect ratio of the parent structures. In this Figure, grey points represent incidences where the parent structures have densities similar to the mean density of the entire Serpens South region; these are small structures that merge at low line brightnesses with the main cloud, and are thus not tracing fragmentation of individual features within the region. Omitting these points, we find a general trend where sibling structures are separated by larger distances when embedded within parent structures with larger aspect ratios, although there is large scatter (r-value of 0.46). 

Figure \ref{fig:frag} (bottom) shows the relationship between the number density of the parent structure and the projected distance separating the embedded structures. The grey points again highlight small structures merging at low line brightnesses with the large-scale \amm\ emission. We further identify incidences where parent structures have aspect ratios $> 4$ (orange). The solid purple line shows the expected separation of structures undergoing Jeans fragmentation at a temperature $T = 11$~K, the mean value found for Serpens South. The lighter purple lines show the expected fragmentation for an isothermal cylinder assuming only thermal support (dashed) and turbulent support (solid), in the case where $R >> H$ and the expected fragmentation length $\lambda_{cyl} = 22H$. Since we find no trend in the \amm\ velocity dispersion with physical scale (see the line width-size discussion in \S\  \ref{sec:sigma_r}), we simply calculate $H_{\sigma_v}$ by setting $\sigma_v$ equal to the mean \amm-derived value over the entire region.  

We see that the expected spherical Jeans fragmentation scale largely provides a minimum separation between sibling structures in Serpens South, but most structures are separated by lengths that are significantly greater, and more consistent with the minimum length scales provided by the filamentary analysis above. This behavior is particularly true for fragmentation in more elongated structures (orange points). We also note that if Serpens South has some inclination with respect to the plane of the sky, the sibling distances will be underestimated by a factor $\cos{i}$, where the true distances will be greater by 1.4 for $i = 45$\arcdeg. Any inclination will therefore move the sibling structure separations further into a regime inconsistent with spherical Jeans fragmentation. \referee{At the spatial scales observed here, the fragmentation of molecular gas within Serpens South is thus not dominated by spherical Jeans fragmentation. Filamentary fragmentation likely plays a greater role, but many structures remain separated by larger distances than the filamentary fragmentation models predict.} 

\begin{figure}[t]
\includegraphics[width=0.48\textwidth]{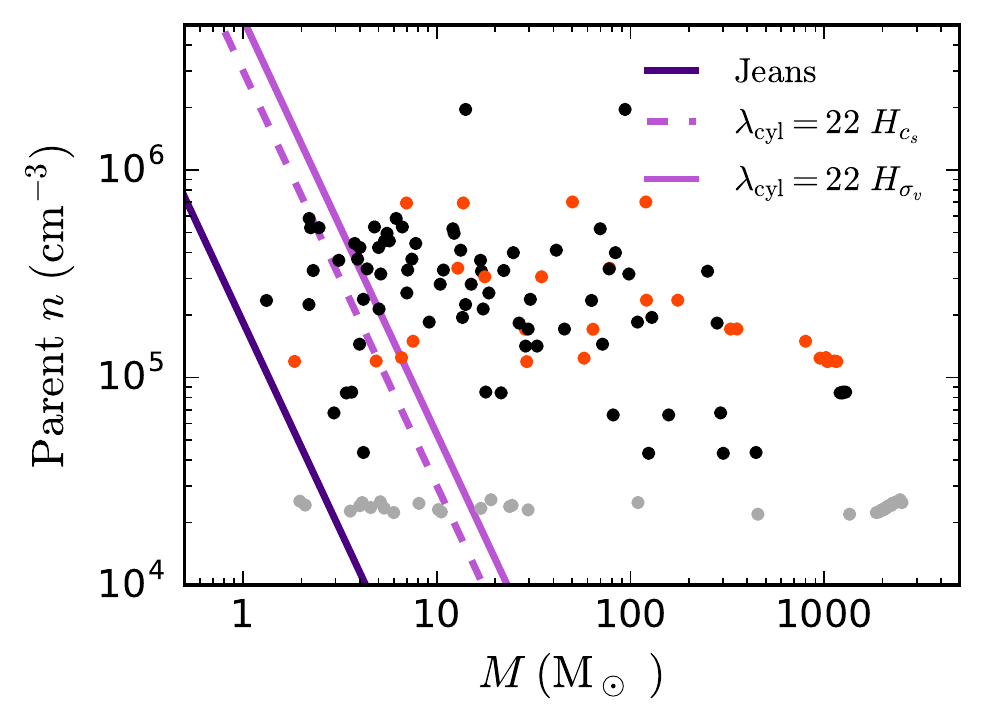}
\caption{Relationship between the mass of sibling structures and the number density of the parent structures. As in Figure \ref{fig:frag}, grey points indicate where the mean density of the parent structure is $\lesssim 2.3 \times 10^{4}~\mathrm{cm}^{-3}$, while orange points highlight incidences where the parent structure is elongated with an aspect ratio greater than 4. The solid purple line shows the expected masses of substructures, given Jeans fragmentation of their parent structure. The lighter purple lines show the expected masses for a fragmenting isothermal cylinder assuming only thermal support (dashed) and turbulent support (solid), where the velocity dispersion for turbulent support is the mean \amm-derived value over the entire region. \label{fig:fragMass}}
\end{figure}

\referee{Similarly to the previous results, we find that the spherical Jeans mass, $M_J$, determined using the observed number density of the parent structure, provides a lower limit to the observed fragment masses.}
This behavior is shown in Figure \ref{fig:fragMass}, where we plot structure mass as a function of parent structure density. As in Figure \ref{fig:frag}, we identify structures that merge at low line brightnesses with the main cloud in gray, and distinguish between structures fragmented from low- and high-aspect ratio structures (black and orange points, respectively). \referee{Most structures have masses significantly greater than the expected Jeans mass of the parent structure at a given density, and there is no clear distinction between fragments embedded within more or less elongated parent structures. }
The objects further down in the dense gas hierarchy tend to be significantly more massive than expected for both thermal Jeans or cylindrical fragmentation. 

 \referee{If the structure masses are overestimated and are instead similar to those reported by \citet{konyves15}, the results shown in Figure \ref{fig:fragMass}\ shift to both lower masses and parent densities for the smaller structures, and some structures become less massive than predicted by a spherical Jeans analysis. Many structures, however, remain significantly more massive than expected by either fragmentation scenario discussed above.} In an analytic model of core growth within filaments, \citet{myers13} shows that more massive cores (of a few M$_\odot$) have been accreting for a longer duration than lower mass cores, and have accreted more gas originally further distant from the filament centre. Evidence of accretion onto the filaments was presented in \papi, and could therefore explain the structure masses that are greater than predicted by simple fragmentation arguments.

A limitation of this analysis is the angular resolution of these data. At the assumed distance to Serpens South, the GBT FWHM subtends 0.07~pc, or approximately the Jeans length at a number density $n \sim 10^5\ \mathrm{cm}^{-3}$ and $T \sim 10$\ K. Figure \ref{fig:frag} (bottom) shows that the densities in Serpens South range from a few $\times 10^4\ \mathrm{cm}^{-3}$ to $\sim 5 \times 10^5\ \mathrm{cm}^{-3}$. At the highest densities, we are unable to resolve the thermal Jeans length, and are likely not capturing additional fragmentation on small physical scales in high density regions, particularly where fragments share similar kinematics. This analysis thus focuses on the fragmentation of the larger-scale filaments in Serpens South. Additional fragmentation of the compact structures is certainly occurring, and can be seen in infrared absorption (see Figure \ref{fig:mm11}) and in combined GBT and Karl G. Jansky Very Large Array data (Friesen et al., in preparation). Since many structures on small scales have small aspect ratios, however (see Figure \ref{fig:aspect}), \referee{and are cold structures containing small non-thermal motions}, we expect fragmentation on smaller scales will be dominated by Jeans fragmentation, in contrast to the fragmentation length scales found here. 

\subsubsection{Structure evolution in non-equilibrium filaments}

\edits{In the previous discussion, we investigated the fragmentation of filaments in Serpens South in an idealized way through comparison with analytic models of infinite, equilibrium cylinders. This analysis thus requires first the formation of relatively smooth filamentary structures, which then fragment into regularly-spaced cores via gravitational instability. In reality, these analytic models likely describe a limiting case of fragmentation in real molecular clouds, where filaments form in a turbulent environment. Here, we discuss the core mass and spacing results in the context of non-equilibrium structures.}

\edits{Filaments may never be equilibrium objects. Colliding-flow simulations of molecular cloud formation suggest that filaments are long-lived features that facilitate the flow of gas from the larger cloud onto star-forming clumps and cores \citep{gomez14}. Dense cores form within the filaments through local instabilities, accreting material from the filaments and collapsing due to the shorter timescale for spherical versus filamentary collapse \citep{burkert04,vazquez07}. In this scenario, cores therefore grow in mass over time as they follow the global gravitational collapse of the filament \citep{gomez14,smith11}. The Serpens South filaments show observational evidence for filamentary accretion and collapse \citep[\papi;][]{kirk13}. Indeed, the fragment masses in Serpens South, particularly those that are greater than expected for thermal or turbulent filamentary fragmentation, may have accreted mass through this mechanism. Over time, the spacing of fragments may also change as they follow the filamentary flow. }

\edits{Furthermore, filaments in molecular clouds are not infinite. The edges of finite filamentary structures can collapse on a timescale shorter than the overall gravitational collapse of an unstable filament, leading to the buildup of dense material at the filament edge \citep{bastien83,burkert04,pon11,toala12}. This sweep-up of material dominates the global gravitational collapse of objects with aspect ratios $> 5$ \citep{pon12}. The largest aspect ratios found in Serpens South through the dendrogram analysis are $\sim 5$, and correspond to the structures within the filament extending north and south of the central protostellar cluster. A dedicated filamentary analysis would likely find a larger aspect ratio for this feature, as several additional structures become kinematically connected to it in the dendrogram analysis, reducing the measured aspect ratio. In Figure \ref{fig:linefits}, several groupings of protostars can be seen at both ends of the filaments that extend north and south of the central cluster, and could have been formed due  to this gravitational focussing effect. This effect may also play a role in regulating the fragment spacing seen in the dense gas.}

\edits{What if filaments and cores form simultaneously? In a turbulent medium, non-linear, turbulence-driven perturbations allow the simultaneous growth of structures over multiple size scales. Some hydrodynamic simulations have indeed shown that filaments and cores grow together, with the initial structure seeded by the cloud turbulence and further developed by self-gravity in the post-shock gas \citep{gong11,gong15}. Filaments and cores can merge and fragment, however, such that the initial structure may not represent the final result. \citet{nakamura14} argue that star formation in Serpens South has been triggered by the collision of filaments, forming the structure seen in dense gas. It is not clear, however, how such an impact would affect the final distribution of dense cores, or whether they would form before, during, or after the collision.}

\section{Summary}
\label{sec:summary}

We have used wide-field mapping of emission from the dense gas tracer, \amm, to examine the structure, stability and fragmentation of dense gas in a young star cluster-forming region, Serpens South. 

\begin{enumerate}

\item
We find that the \amm\ (1,1) emission traces well the submillimeter continuum emission from cold dust. The temperatures derived from the \amm\ analysis are lower overall than those previously derived from dust emission \citep{konyves15}. This difference may be due to the presence of warmer, lower density dust along the line of sight contributing to the observed submillimeter flux density, whereas \amm\ is excited more in the colder, higher density regions. 

\item
We compare the velocity dispersion traced by \amm\ with the thermal sound speed at the \amm-derived gas temperature, finding that the gas traced by \amm\ is sub- or trans-sonic over most of Serpens South. These subsonic regions extend for more than 1\ pc in some regions. Larger, non-thermal motions are visually correlated with the locations of groups of protostars, and may be caused by the impact of outflows on the dense gas or by local gravitational collapse. 

\item
Using a dendrogram analysis, we examined the hierarchical structure of the dense gas. At low \amm\ line brightnesses, we find that most of Serpens South is connected spatially and kinematically. 

\item
We apply the results of the dendrogram analysis to dust continuum-derived H$_2$ column density map, and determine the masses of the hierarchical structures. We find a relationship between structure mass and radius of $M \propto R^{2}$. While the dendrogram analysis is not designed to find filaments, we show that the aspect ratio of structures increases as a function of effective radius, such that most structures with $R_\mathrm{eff} > 0.1$\ pc are significantly more elongated. Where $R_\mathrm{eff} \lesssim 0.1$\ pc, observed structures have aspect ratios of two or less. 

\item
We find that the line width-size relation is flat in Serpens South, such that \amm\ line widths are trans-sonic, on average, over all size scales measured by the extent of our map ($R_\mathrm{eff} \sim 0.9$\ pc). As a result, we find a strong trend of decreasing virial parameter with increasing structure mass. We find extremely low virial parameters for structures on the largest scales probed by our data, suggesting that the previously observed, ordered magnetic field is insufficient to support the region against collapse. This scenario is in agreement with the large radial infall motions previously measured toward some of the filaments. No magnetic field measurements have yet been made specifically in the dense gas, and a more complex magnetic field configuration may be able to support the filaments.

\item
The distances between neighbour structures, embedded within the same parent structure, are significantly greater than expected from a spherical Jeans analysis. We find better agreement with turbulent cylindrical fragmentation models, although with large scatter.  Additional fragmentation beyond the resolution of our data will likely occur on a Jeans scale, due to the small aspect ratios of the most compact observed structures.   

\end{enumerate}

\acknowledgments
We thank the anonymous referee for their clear and detailed comments that improved the paper. RKF is a Dunlap Fellow at the Dunlap Institute for Astronomy \& Astrophysics. The Dunlap Institute is funded through an endowment established by the David Dunlap family and the University of Toronto. The National Radio Astronomy Observatory is a facility of the National Science Foundation operated under cooperative agreement by Associated Universities, Inc. This research made use of astrodendro, a Python package to compute dendrograms of Astronomical data (http://www.dendrograms.org/)

\software{Astropy \citep{astropy}}

\bibliographystyle{apj}

\end{document}